\documentclass[
 aip,
 amsmath,amssymb,
 reprint,
]{revtex4-1}

\usepackage{graphicx}
\usepackage{dcolumn}
\usepackage{bm}

\usepackage[utf8]{inputenc}
\usepackage[T1]{fontenc}
\usepackage{mathptmx}
\usepackage{etoolbox}
\usepackage{subfigure}
\usepackage{amsmath} 
\usepackage{amssymb}
\usepackage{hyperref}
\usepackage[normalem]{ulem}
\usepackage{mathtools} 
\usepackage{comment}
\usepackage{lineno}
\usepackage[table]{xcolor}
\DeclareMathAlphabet{\altmathcal}{OMS}{cmsy}{m}{n}
\usepackage{mathptmx}
\usepackage{booktabs}
\usepackage{xcolor}
\usepackage{amsmath}
\usepackage{soul}
\usepackage{babel}
\usepackage{placeins}
\usepackage{ulem}
\usepackage{booktabs, tabularx}
\usepackage{graphicx}
\usepackage{subcaption} 
\usepackage{subfigure}

\newcommand{\Vsix}{\textcolor{black}}

\makeatletter
\def\@email#1#2{%
 \endgroup
 \patchcmd{\titleblock@produce}
  {\frontmatter@RRAPformat}
  {\frontmatter@RRAPformat{\produce@RRAP{*#1\href{mailto:#2}{#2}}}\frontmatter@RRAPformat}
  {}{}
}%
\makeatother
\begin{document}

\preprint{AIP/123-QED}

\title[A universal relation between intermittency and dissipation within and beyond homogeneous isotropic turbulence]{A universal relation between intermittency and dissipation within and beyond homogeneous isotropic turbulence}

\author{F. H. Schmitt}
\email{felix.schmitt@univ-grenoble-alpes.fr}
\altaffiliation{Univ. Grenoble Alpes, CNRS, Grenoble INP, LEGI, 38000 Grenoble, France}

\author{A. Fuchs}
\email{andre.fuchs@uni-oldenburg.de}
\affiliation{Institut für Physik und ForWind, Universität Oldenburg, Küpkersweg 70, 26129 Oldenburg, Germany}

\author{J. Peinke}
\email{joachim.peinke@uni-oldenburg.de}
\affiliation{Institut für Physik und ForWind, Universität Oldenburg, Küpkersweg 70, 26129 Oldenburg, Germany}

\author{M. Obligado}
\email{martin.obligado@centralelille.fr}
\affiliation{Univ. Lille, CNRS, ONERA, Arts et Métiers ParisTech, Centrale Lille, FRE 2017 - LMFL - Laboratoire de Mécanique des fluides de Lille - Kampé de Feriet, F-59000 Lille, France}

\date{\today}

\begin{abstract}
Fundamental quantities of turbulent flows, such as the dissipation \Vsix{parameter} $C_\varepsilon$ and the intermittency \Vsix{parameter} $\mu$, are examined in relation to each other for a broad class of inhomogeneous turbulent flows. In the context of the energy cascade, it is known that $C_\varepsilon$ reflects its basic overall properties, while $\mu$ quantifies the intermittency \Vsix{of} the cascade. Using an extensive hot-wire dataset of turbulent wakes, grid-generated turbulence, and an axisymmetric jet, we individually analyze these quantities as one-dimensional surrogates of the energy cascade, considering only data that exhibit consistent scaling behavior. We find that $\mu$ is inversely proportional to $C_\varepsilon$, offering a new empirical principle that bridges the gap between large and small scales in arbitrary turbulent flows. The generalized framework presented here recovers the established values and trends reported for homogeneous isotropic turbulence, and expands them to cover inhomogeneous situations.
\end{abstract}

\maketitle

Despite recent advances in the study of turbulent flows~\cite{sreenivasan2025turbulence}, solving the Navier–Stokes equations at large Reynolds numbers and for an arbitrary set of boundary conditions still remains beyond current numerical and theoretical capabilities~\cite{frisch1995turbulence, alberti2023still}. The lack of a simplifying framework for complex turbulent flows hinders modeling efforts across various fields, as there are many engineering problems where turbulence plays a crucial role, turbulent wakes generated by wind turbines being among the most relevant industry-related applications~\cite{stevens2017flow, neunaber2020distinct}. The most common approach to tackle turbulence is to define equations in which factors or exponents are determined by dimensional arguments or experiments. Within this approach, a relevant question that arises concerns the universality of these \Vsix{parameters}, even when considering statistically steady homogeneous isotropic turbulence (HIT) only~\cite{praskovsky1994measurements, renner2002universality}.

Since the publication of Kolmogorov's phenomenology in 1941 (K41)~\cite{kolmogorov1941local}, there has been intense work to shed light on this question. This was enhanced by the publication of Kolmogorov's revised theory in 1962~\cite{kolmogorov1962refinement} and further developments like the multifractal formalism ($cf.$ Sreenivasan's review on the topic~\cite{sreenivasan1991fractals}), as further \Vsix{parameters} appeared. Although there were some deviations in experimental results, the overall opinion -- at least until the 1990s -- was that such \Vsix{parameters}, for sufficiently large Reynolds numbers, are universal for fully developed turbulence, which led to a focus on their numerical values~\cite{praskovsky1994measurements, sreenivasan1995universality, arneodo1996structure, praskovsky1997comprehensive, yeung1997universality}.

Among all relevant parameters, two of the most important \Vsix{ones} to describe turbulent flows are, arguably, the dissipation \Vsix{parameter} $C_\varepsilon$ \Vsix{(also known as the dissipation constant)} and the intermittency \Vsix{parameter} $\mu$. The former has been identified as one of the most important parameters to describe turbulence~\cite{lumley1992some}, as it can be directly linked to the energy cascade~\cite{vassilicos2015dissipation}. On the other hand, the \Vsix{parameter} $\mu$, first introduced by Kolmogorov's 1962 theory~\cite{kolmogorov1962refinement}, and later redefined by Castaing and collaborators~\cite{castaing1990velocity, arneodo1996structure}, quantifies departures from self-similarity within the inertial range. \Vsix{Despite its relevance, a physical interpretation of $\mu$ as well as a direct derivation from the Navier--Stokes equations is still missing.} Nevertheless, and independently on the formalism used to interpret this quantity, it provides a scalar value that quantifies the \Vsix{intensity} of intermittency as departures from Gaussianity of \Vsix{the velocity} increment \Vsix{probability density functions (PDFs)} in the inertial range. \Vsix{Note that there exists at least three distinct definitions of intermittency in turbulence research. The first refers to intermittent changes of the turbulent and non- (or less) turbulent flow packages, typically seen at the border of turbulent regimes. Such intermittent switches lead to non-Gaussian distributions of velocity increments. This is also termed intermittency, \emph{e.g.}~\cite{zhou2023appearance}, although it pertains to scales outside the inertial range. The second refers to the non-Gaussianity of velocity increment statistics at a single scale only, \emph{e.g.} at the limit of the inertial range toward small scales. The third definition, following Kolmogorov and Castaing~\cite{kolmogorov1962refinement,castaing1990velocity}, is adopted in this work and concerns the change from Gaussian large scale velocity increment statistics to non-Gaussian statistics at smaller scales.}

Over the last two decades, the universality of $C_{\varepsilon}$ (sometimes referred to as ``zeroth law''  of turbulence~\cite{sreenivasan2025turbulence}) has been challenged by several experimental and numerical studies~\cite{vassilicos2015dissipation}.

The dissipation \Vsix{parameter} is defined as 
\begin{equation}
    C_{\varepsilon} = \frac {\varepsilon \: L} {u^{\prime 3}},
    \label{dissipation_constant}
\end{equation}

\noindent with $\varepsilon$ the mean turbulent kinetic energy dissipation rate \Vsix{(often denoted as $\overline{\varepsilon}$)}, $u^{\prime}$ the \Vsix{root mean square} value of the streamwise fluctuating velocity and $L$ the integral length scale. The importance of $C_\varepsilon$ for modeling arises from the fact that it gives an estimation of the mean dissipation rate, which is often needed to have a closed system of equations~\cite{george1989self}. According to K41 phenomenology, considering steady, homogeneous turbulence, $C_\varepsilon$ should depend on boundary conditions, being constant for sufficiently large values of the Taylor-length based Reynolds number $Re_\lambda = u^{\prime} \, \lambda / \nu$, with $\lambda$ the Taylor length scale and $\nu$ the kinematic viscosity of the fluid. 

In several free-shear flows, results not consistent with K41 phenomenology have been reported, and the value of $C_\varepsilon$ at the centerline has been found to decrease with increasing $Re_\lambda$~\cite{vassilicos2015dissipation,ortiz2021high}. More precisely, for this so-called non-equilibrium cascade, $C_\varepsilon$ is expected to be linear with $\sqrt{Re_G}/Re_\lambda$ (with $Re_G$ a Reynolds number that depends on the inlet properties of the flow, such as the wind tunnel inflow velocity and the diameter of the wake-generating object). Most of the work on this topic aims at quantifying the dependence of $C_\varepsilon$ over a broad range of Reynolds numbers. These studies concern laminar inflows, and little attention has been paid to further details of the energy cascade, \Vsix{such as intermittency}.

Intermittency is one of the causes of deviations from the K41 phenomenology. One of its most significant features is that velocity increments, $u_r(x) := u(x+r) - u(x)$, exhibit extreme events at small scales $r$. There are two standard approaches to quantify intermittency.

The most common one is to investigate the scaling behavior of generalized structure functions \Vsix{$S_n(r) := \overline{{u_r(x)}^n}$ of a given order $n$, where the over-bar implies a spatial averaging.} As K41 implies that $\overline{u_r(x)^n} \propto (\varepsilon \: r)^{\zeta_n}$, with the exponent $\zeta_n$ equal to $n/3$, many works focus on corrections either to the power-law form or to the value of the exponent \cite{sinhuber2017dissipative, reinke2018universal} ($cf.$~\cite{sreenivasan1997phenomenology}).
This approach has been extensively used for HIT, resulting in various scaling models for the exponent $\zeta_n$ of the structure functions. Starting from Kolmogorov’s 1962 theory (based on a lognormal model for the two-point energy dissipation), many improvements and alternative frameworks have been proposed, such as the works of She-Leveque~\cite{she1994universal}, Castaing-Yakhot~\cite{castaing1996temperature, yakhot1998probability}, and L'vov-Procaccia~\cite{l2000analytic}. For a recent review, the reader is referred to Benzi \& Toschi~\cite{benzi2023lectures}.

In this structure function approach, two questions arise. On the one hand, as discussed above, an open question is how the higher-order exponents $\zeta_n$ deviate from the relation $\zeta_n = n/3$ implied by K41. Typically, these deviations become significant for $n > 4$, while differences between \Vsix{distinct} anomalous scaling models can be detected for $n > 6$. On the other hand, another open discussion concerns how deviations from the expected scaling of the structure function can be understood in the context of non-homogeneous turbulent flows~\cite{Arad1998,biferale2002probing,kurien2002measures,benzi2023lectures,zhou2021turbulence}.

A second approach is to investigate directly the form of the PDFs \Vsix{$p(u_r(x)/\sigma_r)$ at different values of $r$, where $\sigma_r$ denotes the standard deviation of $u_r(x)$. (Note that we refer here to the shape of the PDF, as its width is removed by normalization with 
$\sigma_r$, which is directly related to the power spectrum.) The form is captured via the scale-dependent shape parameter $\Lambda^2(r)$ (often denoted as $\lambda^2(r)$, but to avoid confusion with the Taylor length scale, we use the capital Greek letter). This quantity is linked to the variance of the distribution of ln($\varepsilon_r(x)$), with ln the natural logarithm of the -- also scale-dependent -- energy dissipation $\varepsilon_r(x)$~\cite{kolmogorov1962refinement,castaing1990velocity}. It can be estimated following Chilla \emph{et al.}~\cite{chilla1996multiplicative} as,}

\begin{equation}
    \Lambda^2(r) = \frac {\ln \left(F(u_r(x))/3 \right)} {4},
    \label{lambda}
\end{equation}

\noindent \Vsix{where $F$ denotes the flatness, such that for every scale $r$ from the largest to the smallest, a shape parameter is obtained. After Kolmogorov~\cite{kolmogorov1962refinement} and Castaing~\cite{castaing1990velocity} the scale dependence in the inertial range, for $\eta \ll r \ll L$ (with $\eta = (\nu^3 / \varepsilon)^{1/4}$, being the Kolmogorov length scale), is expected to follow the relation,}

\begin{equation}
    {\Lambda^2(r) = \Lambda_0^2 + \mu/9 \; \ln(L/r),}
    \label{mu}
\end{equation}

\begin{figure}[h!] 
    \centering    \includegraphics[width=\linewidth]{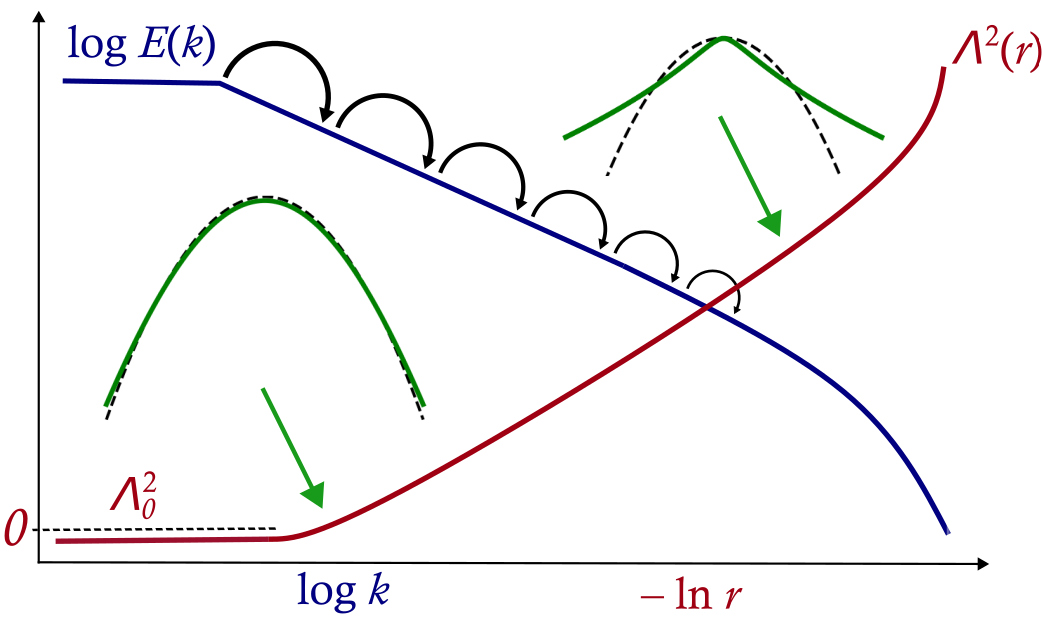} 
    \caption{Schematic representation of the spatial expansion of the turbulence cascade: the blue curve shows the energy spectrum, with energy transported to smaller scales, while the red curve illustrates the evolution of the shape \Vsix{parameter} $\Lambda^2(r)$ for normalized velocity increments. Note that in this sketch, $\Lambda_0^2$ is approximately zero.}
    \label{Turbulence_scheme}
\end{figure}

\noindent \Vsix{with $\Lambda_0^2$ a parameter that depends on the properties of the flow (see also figure~\ref{Turbulence_scheme}). Note that the relation~(\ref{mu}) is consistent with the power-law scaling behavior of the energy spectral density in the inertial range as well as the structure functions, as discussed in~\cite{castaing1990velocity} (see eq.~(\ref{energy_spectrum})). The key assumptions of Kolmogorov regarding eq.~(\ref{mu}) are HIT, high $Re_\lambda$ and that energy is transported across scales in a cascade process, whereby it is transferred stochastically from larger to smaller scales. However and despite its original assumptions, eq.~(\ref{mu}) has been shown to remain valid for inhomogeneous flows~\cite{morales2012characterization}, and even appears to capture key features of cascade-like processes in financial time series~\cite{ghashghaie1996turbulent}. As a remark, the difference between $\mu$ and $\zeta_4$, the fourth-order structure function exponent within the inertial range should be pointed out. In general, the fourth-order structure function $S_4(r)$ and the shape parameter $\Lambda^2(r)$ are directly comparable. However, since the kurtosis is the fourth-order moment normalized by the variance, $\Lambda^2(r)$, in contrast to $S_4(r)$, serves as a measure of the form of the PDFs and their deviation from Gaussianity, as will be shown later. In this sense, $\Lambda^2(r)$ captures all higher order moments or structure functions.}

Within this rationale, the \Vsix{ evolution of} non-Gaussianity of the velocity increment distributions \Vsix{over the scale $r$} can be quantified by \Vsix{the} single scalar parameter \Vsix{$\mu$}. The value of $\mu$ has been commonly accepted to be constant at large values of $Re_\lambda$. The reported values range between $0.2$ and $0.4$ for HIT so that a value of $0.26$ has been established as representative of turbulent flows~\cite{arneodo1996structure,vinnes2023characterizing,neunaber2020distinct}. Outside HIT, several works report values substantially larger than 0.4. For instance, Gibson $et \, al.$ estimated $\mu \approx 0.5$ in measurements over the open ocean surface~\cite{gibson1970measurements}.

Despite intense research on turbulent flows, no relation between $\mu$ and $C_\varepsilon$ has been predicted or found. Even for the case of statistically steady HIT flows, where these \Vsix{parameters} are expected to reach constant values at large $Re_\lambda$, their relation for a given set of boundary conditions remains unexplored. Furthermore, the universality of these \Vsix{parameters} when flows depart from statistically steady HIT is still an open question, with a growing amount of evidence pointing against this feature for several inhomogeneous and/or statistically unsteady flows. While most studies deal with homogeneous turbulence, in this study we aim to extend the investigation to inhomogeneous turbulent flows. These flows present several challenges when defining both $C_\varepsilon$ and $\mu$, and therefore require a careful data selection process (detailed in section~\ref{Res}). We therefore do not claim that the large body of work performed previously has to be revisited, but rather show how the well-established properties of these \Vsix{parameters} are modified when inhomogeneity is taken into account. Hence, we present a detailed experimental study on turbulent wakes, supported by previously published datasets on grid-generated turbulence and a turbulent axisymmetric jet. Our work therefore shows how previous results remain valid but can be generalized through a novel relation between these \Vsix{parameters}.

\begin{figure} \centering
    \includegraphics[width=\linewidth]{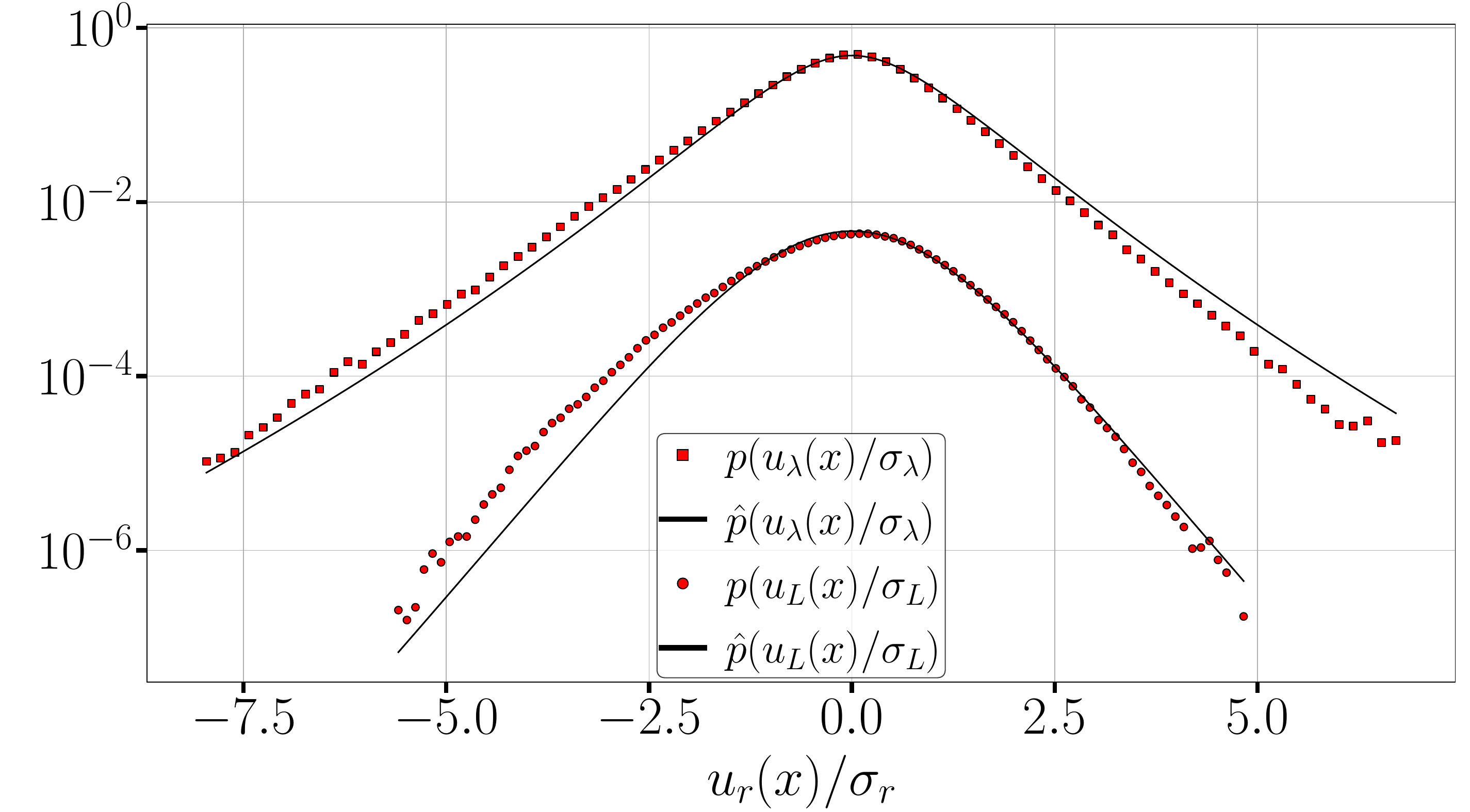} 
    \caption{\Vsix{Exemplary PDFs of velocity increments $p(u_r(x)/\sigma_r)$ for one velocity time series within the highly restricted dataset. In blue, the velocity increments for the integral length scale and in red for the Taylor length scale are presented. The shape \Vsix{parameter} at the integral length scale has a value of $0.043$ and at Taylor length scale a value of $0.125$. Additionally, for each of those two values of $r$, a Castaing curve~\cite{castaing1990velocity, morales2012characterization, neunaber2019stochastic} $\hat{p}(u_r(x)/\sigma_r)$ in black is plotted over the experimental data. For the Castaing curve, the used value of $\Lambda^2$ corresponds to the extracted value from the experimental data. The mean squared deviation for the blue and red data from its Castaing curve is $0.067$ and $0.069$, respectively. For reasons of visibility, the curves are shifted vertically.}}
    \label{PDF_example}
\end{figure}

To analyze such non-HIT data, we use eq.~(\ref{dissipation_constant}) for $C_{\varepsilon}$ and eq.~(\ref{lambda}) and eq.~(\ref{mu}) for $\mu$. For $C_{\varepsilon}$, we justify our approach by considering each velocity time series as a 1$\,$D surrogate. This has been done before, for example, for fractal-grid turbulence~\cite{vassilicos2015dissipation} and turbulent boundary layer flows~\cite{nedic2017dissipation}. For the estimation of $\mu$, \Vsix{we justify this approach based on the quality of the PDFs of velocity increments, $p(u_r(x)/\sigma_r)$, in comparison to the Castaing curves $\hat{p}(u_r(x)/\sigma_r)$ as shown in figure~\ref{PDF_example}. The quality of the fits was quantified as follows. For each velocity time signal, two empirical PDF of velocity increments for different scales $r$ were evaluated. To ensure the significance, all bins with less than 10 entries were removed. The chosen scales were the integral length scale $L$ and the Taylor length scale $\lambda$. The mean squared deviations between each empirical PDF of the velocity increments and the corresponding Castaing curve ${\hat p(u_r(x)/\sigma_{r})}$ were calculated, obtaining the errors ${e_c(L)}$ and ${e_c(\lambda)}$. The overall deviation per velocity time series $\overline{e_c}$ is then calculated by computing the average of ${e_c(L)}$ and ${e_c(\lambda)}$. The Castaing curve can be obtained by using the eqs.~(\ref{castaing_curve}) and~(\ref{castaing_curve_sigma})~\cite{castaing1990velocity, morales2012characterization, neunaber2019stochastic} in the appendix~\cite{SM}.} 

Our approach aims to extend our understanding of HIT flow conditions. In particular, we aim to quantify the intermittency of such flows in a general way, independently of the turbulence model used to interpret the results. Therefore, a good fit of $p(u_r(x)/\sigma_r)$ is sufficient for our objectives of having a scalar surrogate for the internal intermittency of the flow. It is important to mention that the refined estimation of $\mu$ used in this work does not require any assumption regarding the behavior of the structure functions.

The manuscript is structured as follows: the experimental setup and the different turbulent flows considered in this work are described in Sec.~\ref{ES}. The main results, including the relation between $\mu$ and $C_\varepsilon$, are presented in Sec.~\ref{Res}. These findings and their implications for inhomogeneous turbulence are further discussed in Sec.~\ref{Dis}. Additional analyses, robustness checks, and methodological details are provided in the appendix~\cite{SM}.

\section{Experimental setup \label{ES}}

We conducted wind tunnel experiments in LEGI lab, Grenoble, by generating a turbulent wake behind porous and impermeable objects under different inflow conditions. The facility is a closed-circuit system with test section dimensions of 0.75 $\times$ 0.75 $\times$ $4\,$ m$^3$. The objects were cylinders and disks while both the diameter and solidity vary (see table~\ref{PhD measurements in LEGI 2023} in the appendix~\cite{SM}). To reduce the blockage, the cylinders were directly screwed to the walls of the tunnel while the disks were hold at the tunnel's centerline with four thin piano wires which went through taped holes in the tunnel walls. For varying the inflow, both the entrance of the test section and the tunnel's freestream velocity were modified. We used laminar inflow, and different turbulent background flows generated by a static regular grid with cylindrical bars or an active grid driven in a triple random mode (information about the turbulent flow produced by grids is reported in a previous work~\cite{ferran2023experimental}). The wake of the objects was scanned by a $1\,$D hot wire with a temporal resolution of $50\,$kHz (with an anti-aliasing filter set at $30\,$kHz) which was traversed in streamwise and spanwise directions. The conversion from the temporal to the spatial domain was made using Taylor's hypothesis. The nearest streamwise position to the wake generator was $5\,d$ and the farthest was $200\,d$, with $d$ being the diameter of the object. \Vsix{Additionally, the background flows without the object were also measured.} For every position the data was acquired for $120\,$s. In total, $15$ configurations were carried out. \Vsix{For simplicity, only points in the vicinity of the centerline are considered, resulting in a total of 498 data points forming the initial dataset prior to the selection process. Note that figure~\ref{Wind_tunnel_Grenoble} in the appendix~\cite{SM} completes the description concerning turbulent wakes, by showing a scheme of the experimental setup in LEGI, Grenoble.}

\Vsix{In addition to the experimental dataset on turbulent wakes, our study is extended by means of $110$ velocity time series from three previously published experimental datasets: 1) decaying passive-grid generated turbulence within the Boundary Layer Wind Tunnel from the Laboratoire de Mécanique des Fluides in Lille (marked by G20, G21, G22). The measurement points are located between $3.3 \, M$ and $54 \, M$, using $M$ as the mesh size. For further details, $cf.$~\cite{ferran2023characterising}. 2) Decaying passive-grid generated turbulence within the Lespinard Wind Tunnel from LEGI in Grenoble (marked by G23, G24). The measurement points are located between $1 \, M$ and $30 \, M$, using $M$ as the mesh size. For further details, $cf.$~\cite{mora2019energy}. 3) An axisymmetric turbulent free jet (marked by J25). The measurement point is located at $125 \, d$, using $d$ as the nozzle diameter. For further details, $cf.$~\cite{renner2001experimental}. For all added datasets, data acquisition was also achieved by using a $1\,$D hot wire. The two studies investigating passive-grid turbulence were selected as one decays in agreement with Kolmogorov's dissipation scalings~\cite{kolmogorov1941local} while the other study shows a decay with non-equilibrium behavior~\cite{vassilicos2015dissipation}. The study of the jet was selected because it completes the series of canonical free-shear flows at steady state. Table~\ref{PhD measurements in LEGI 2023} in the appendix~\cite{SM} shows the entire dataset used in this paper and how different cases are marked as well as some further relevant details. Preliminary checks suggest that all results reported in this work remain valid for large-shear non-centerline data. }

\section{Results \label{Res}}

For all those points from all mentioned datasets, a rigorous analysis of common $1\,$D-free-shear turbulence \Vsix{parameters} is derived and checked individually. Only time series that have the following characteristics were kept for further analysis: 

\noindent a) the turbulence intensity $TI = u^{\prime} / \overline{u}$ with $\overline{u}$ being the actual mean velocity, decreases in streamwise direction,
\noindent b) the large scale increments have an almost Gaussian distribution ($\Lambda_{0}^{2} < 0.025$)
\noindent c) the slope of the energy spectrum within the inertial range is smaller than $-1.25$,
\noindent d) the length of the inertial range for the energy spectrum is larger than $0.5$ decades,  
\noindent e) the $R^2$-value of the linear regression of the fit of $\Lambda^{2}$ is larger than $0.99$, 
\noindent f) the mean squared deviation from the calculation of $\Lambda^{2}$ is in average smaller than $0.11$, 
\noindent g) the flatness of the velocity time signal is around $3$ ($2.67 < F_u < 3.1$) and  
\noindent h) the skewness of the time signal is below $0.1$. Details and an example of the analysis is shown and discussed in the appendix~\cite{SM}.

After this conditioning, 312 velocity time series remained for the analysis. Each of them represents at least $2\times10^4$ integral time scales.

The integral length scale $L$ is calculated by integrating the autocorrelation function until it crosses $1/\mathrm{e}$ ($\mathrm{e}$ being the Eulerian number). This estimation is known to provide a smoother dependence of $L$ on the governing parameters. While not integrating up to the first zero crossing results in smaller values of $L$ (and therefore of $C_\varepsilon$), all trends are preserved \cite{mora2020estimating}. Other integration limits were tested, and the trends presented in this Letter remained unchanged. $\varepsilon$ is derived -- the common definition for HIT conditions is used -- by integrating the dissipation spectrum $\varepsilon = \int_{0}^\infty  15 \nu \: k^2 E(k) \: \mathrm{d} k$ (with $k$ the wavenumber in the Fourier space) and modeling the high-frequency noise with an extrapolated fit (a test with different fitting limits and functions gave an estimated error in the estimation of $\varepsilon$ in the order of $5\,\%$). 

\begin{figure}[h!] 
    \centering    \includegraphics[width=\linewidth]{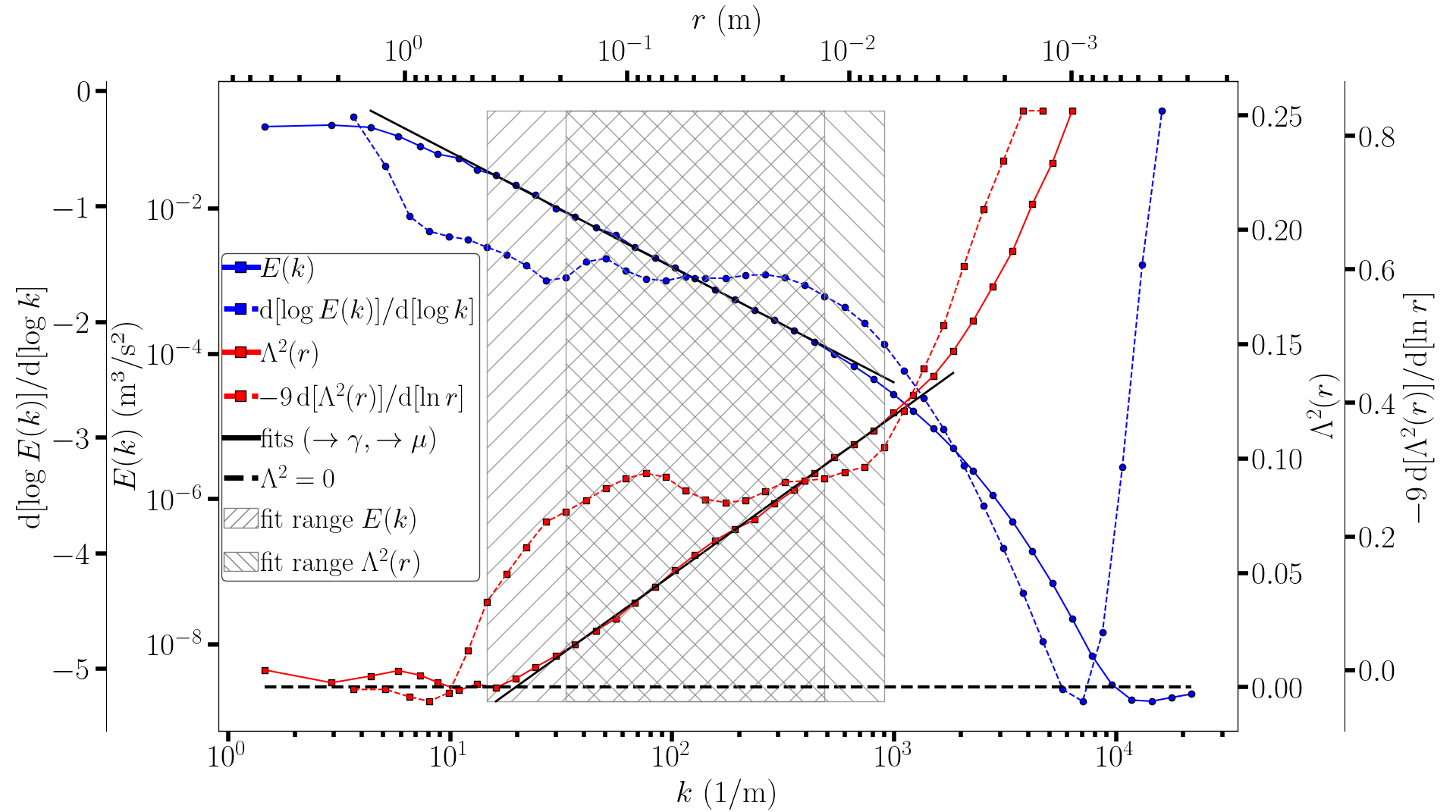} 
    \caption{\Vsix{The evolvement of the energy spectral density $E(k)$ and the shape \Vsix{parameter} $\Lambda^2(r)$ are plotted over the scales $k$ and $r$, respectively. Additionally, the derivatives of the quantities in log/log and lin/log, respectively, are presented. Note that the axis correspond to the derivative of $\Lambda^2(r)$ already contains the prefactor 9, so that the value of $\mu$ can be read directly. The fit areas are marked by hatched rectangles while the fits itself are visualized by solid black lines. Additionally, the value of $\Lambda^2 = 0$ is indicated with a black dashed line. Note that $\Lambda^2$ is almost $0$ for very large increments.}}
    \label{scheme_data_example}
\end{figure}

We extracted, after eq.~(\ref{mu}), $\mu$ by fitting the evolution of the shape \Vsix{parameter} $\Lambda^{2}$ within the inertial range~\cite{castaing1990velocity}. \Vsix{Figure~\ref{scheme_data_example} shows, similar to the scheme in figure~\ref{Turbulence_scheme}, an example of the analysis using a velocity time signal from our wake experiment. The example velocity time signal belongs to the case D15 (see table~\ref{PhD measurements in LEGI 2023} in the appendix~\cite{SM}). Both the energy spectral density and the shape parameter have a clear power-law and logarithmic behavior, respectively, over the inertial range. Further details for this particular velocity time series are shown in table~\ref{data_example} in the appendix~\cite{SM}. To determine the inertial range in $r$-space, the derivative of $\Lambda^2(r)$ in lin/log-representation was used, where an algorithm evaluates the limits of a plateau within the derivative. These limits are used as the fit bounds. The dependence of the extent of the inertial range on $\mu$ is also shown in the appendix~\cite{SM} in figure~\ref{mu_L_div_eta}. Note that in figure~\ref{scheme_data_example}, the shape parameter on large scales is close to zero.}

\Vsix{Moreover, the parameters governing the spectral law within the inertial range were also evaluated. Following the notation of \Vsix{Mydlarski \& Warhaft}~\cite{mydlarski1996onset},}

\begin{equation} 
    \Vsix{
    {E(k)=C_k \:\: \varepsilon^{2/3} \: k^{-5/3} \: (k \: \eta / 2 \: \pi)^{-\gamma + 5/3},}
    \label{energy_spectrum}
    }
\end{equation}

\noindent \Vsix{the energy spectral density should exhibit a power-law behavior within the inertial range, where $C_k$ is the Kolmogorov parameter (also known as the Kolmogorov constant) and $\gamma$ is the absolute slope of the energy spectrum. The K41 prediction for the energy spectral density within the inertial range is complemented with the correction $ (k \: \eta / 2 \: \pi)^{-\gamma + 5/3}$. K41 is therefore recovered when $\gamma=5/3$ and, for large values of $Re_\lambda$, $C_k$ is a constant (commonly accepted to be around 0.5 for HIT~\cite{sreenivasan1995universality}). The value of 5/3 emerges through a dimensional argument which followed the first two similarity hypotheses of Kolmogorov. Over the last years, several works reported values of $\gamma$ that deviate from the expected 5/3 value~\cite{steiros2022balanced, rodriguez2023not, neunaber2020distinct}, even at large values of $Re_\lambda$, most likely due to intermittency corrections. On the other hand, less evidence of deviations from $C_k \, \approx \, 0.5$ has been reported in the literature~\cite{meyers2008functional}).}

\Vsix{The estimation of $C_k$ and $\gamma$ was done similar to the estimation of $\mu$, by analyzing the derivative of $E(k)$ in log/log-representation in $k$-space, as can be seen in figure~\ref{scheme_data_example}. For all velocity time series, the fits were checked individually. Furthermore, a sensitivity analysis was performed to see how different versions of the algorithm detect different plateaus. No systematic effects have been found. In figure~\ref{scheme_data_example}, the inertial ranges in $r$-space and $k$-space clearly overlap, which was observed for all velocity time series. However, there was also always a shift, such that the inertial range of the shape parameter is more to smaller scales, as visible in figure~\ref{scheme_data_example}. No parameter was found that explained the distinct shift (note that our conversion from $k$-space to $r$-space was performed by $r=2\pi/k$).}

\begin{figure}
    \centering    \includegraphics[width=\linewidth]{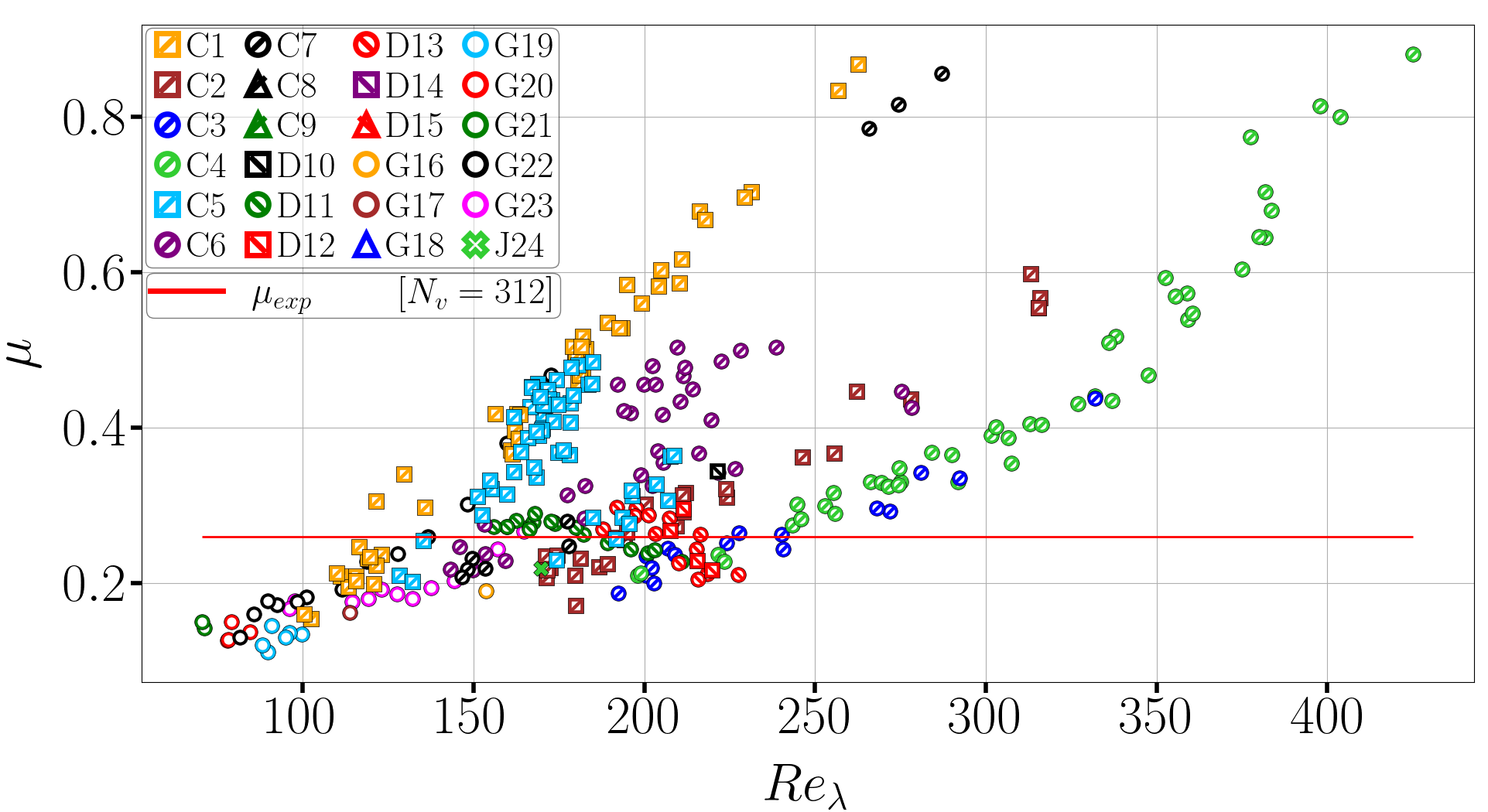} 
    \caption{\Vsix{$\mu$ as a function of $Re_\lambda$ for the highly restricted data. The red line corresponds to a commonly accepted value for $\mu$ for homogeneous isotropic turbulence~\cite{arneodo1996structure}. In general, the symbols in the legend are identical for all figures throughout this manuscript and correspond to the configurations in table~\ref{PhD measurements in LEGI 2023} in the appendix~\cite{SM}. For laminar inflow, squared markers are used. For the regular grid and the active grid, markers are shaped as circles and triangles, respectively. For cylinders as generators, the markers contains a ``$/$'' (case names start with ``C'') while for disks a ``$\backslash$'' (case names start with ``D'') is used. Hollow circular markers means no object and is equivalent to grid turbulence (case names start with ``G''). The ``x'' marker indicates a free jet (case names start with ``J''). $N_v$ indicates the number of velocity time series shown in this plot.}}
    \label{mu_reynolds_num}
\end{figure}

First, we investigate if the parameters $\mu$ and $C_\varepsilon$ present a constant value in our dataset, including all the flows studied. In figure~\ref{mu_reynolds_num} it is clear that $\mu$ presents a large spreading and it does not approach towards an asymptotic limit at large values of $Re_\lambda$. 

\begin{figure}
    \centering    \includegraphics[width=\linewidth]{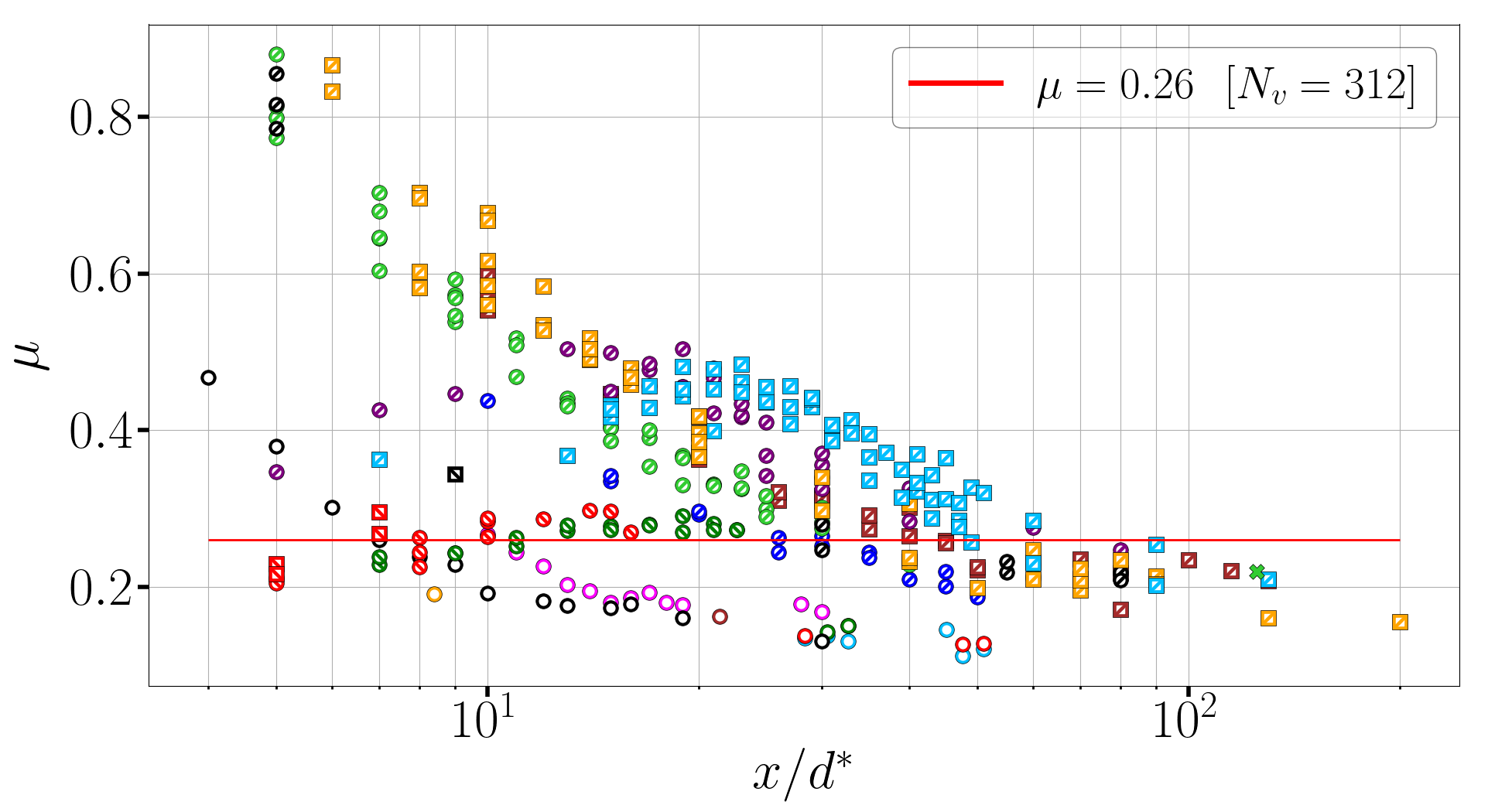} 
    \caption{\Vsix{$\mu$ as a function of $x/d^*$, the non-dimensional streamwise distance to the turbulence generator for the highly restricted dataset. For the configurations with the disks and cylinders, $d^*$ is equal to the object diameter. For grid-generated turbulence, $d^*$ is equal to mesh size and for the jet, the diameter of the nozzle is taken. The red line indicates a commonly accepted value for $\mu$ for homogeneous isotropic turbulence~\cite{arneodo1996structure}. The symbols and corresponding configurations are shown and explained in figure~\ref{mu_reynolds_num} and table~\ref{PhD measurements in LEGI 2023} in the appendix~\cite{SM}. $N_v$ indicates the number of velocity time series shown in this plot.}}
    \label{object_distance}
\end{figure}

\Vsix{To identify where the high $\mu$ values occur, figure~\ref{object_distance} shows the distribution of $\mu$ in dependence on the normalized streamwise distance to the turbulence-generating object from the perspective of the hot-wire probe. The streamwise distance $x$ is normalized by $d^*$, a typical characteristic length of the turbulence-generating object. For wakes, it is the diameter of the disk or cylinder while for grid-turbulence it is the mesh size and for the jet the nozzle diameter is used. For the shown highly restricted data, $x/d^*$ has a range between $4$ and $200$.}

\begin{figure}
    \centering    \includegraphics[width=\linewidth]{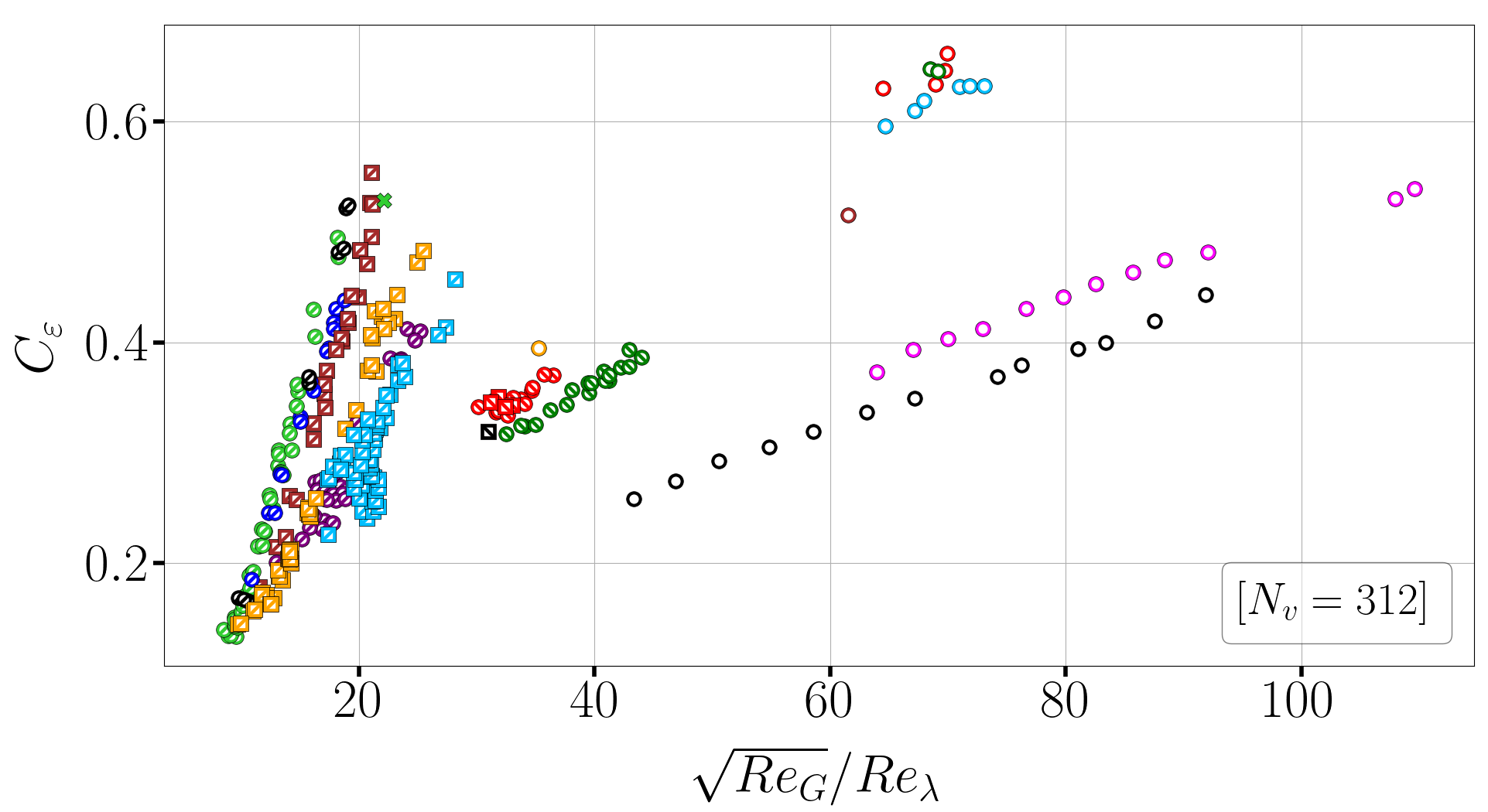} 
    \caption{\Vsix{$C_\varepsilon$ versus $\sqrt{Re_G}/Re_\lambda$ for the highly restricted data. The symbols and corresponding configurations are shown and explained in figure~\ref{mu_reynolds_num} and table~\ref{PhD measurements in LEGI 2023} in the appendix~\cite{SM}. $N_v$ indicates the number of velocity time series shown in this plot.}}
    \label{c_epsi_global}
\end{figure}

Figure~\ref{c_epsi_global} shows that $C_\varepsilon$ also behaves similarly, $e.g.$ $C_\varepsilon$ does not have a constant value. Having a closer look on the data we see that for each set of boundary conditions, a linear relation between $C_\varepsilon $ and $\sqrt{Re_G}/Re_\lambda$ is consistent with the dissipation scalings from non-equilibrium turbulence~\cite{vassilicos2015dissipation}. Remarkably, when such conditions are changed, the linear relation still applies but with a different slope, showing the emergence of new length scales that affect the dissipation scalings. \Vsix{In figure~\ref{skewness_cepsi_e} in the appendix~\cite{SM}, we additionally provide the relation between the skewness $S_u$ of the velocity time series over $C_\varepsilon$ for the highly restricted dataset. Due to our data selection, where we are not restricting ourselves to HIT data, the values of $S_u$ are both positive and negative. Furthermore it can be seen that there is no clear trends with $C_\varepsilon$, thus based on our data, no conclusive result on the skewness of the velocity time signals are obtained.}

\begin{figure}
    \centering    \includegraphics[width=\linewidth]{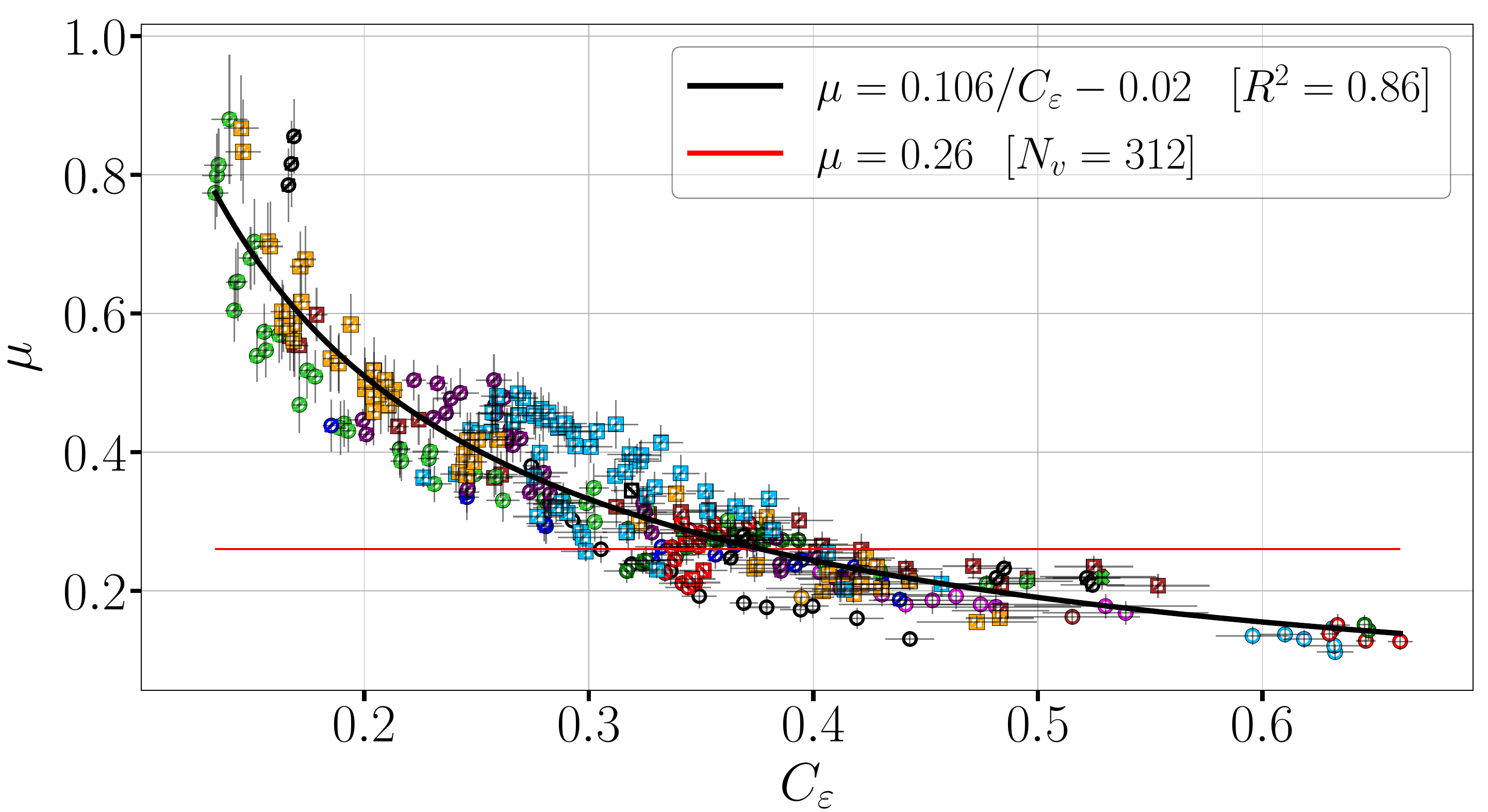} 
    \caption{\Vsix{$\mu$ as a function of $C_\varepsilon$ for the highly restricted data. The red line indicates a commonly accepted value for $\mu$ for homogeneous isotropic turbulence~\cite{arneodo1996structure}, and the black line corresponds to a least-square fit ($R^2=0.86$ with the fitted values of $0.106 \pm 0.002$ and $0.022 \pm 0.009$). The error bars indicate the uncertainties for both $\mu$ and $C_\varepsilon$. The symbols and corresponding configurations are shown and explained in figure~\ref{mu_reynolds_num} and table~\ref{PhD measurements in LEGI 2023} in the appendix~\cite{SM}. $N_v$ indicates the number of velocity time series shown in this plot.}}
    \label{mu_c_epsi_e_error}
\end{figure}

Overall, our dataset does not show a universal or predictable behavior in terms of neither $C_\varepsilon$ nor $\mu$. Next, we want to see if there is a relation between $\mu$ and $C_\varepsilon$. Figure~\ref{mu_c_epsi_e_error} shows that $\mu \propto 1/C_\varepsilon$, or, respectively, that the product of those two quantities is constant. \Vsix{For $C_\varepsilon$, the uncertainty is estimated based on the added fraction of $\varepsilon$ when modeling the high-frequency noise with an extrapolated fit. For $\mu$ the uncertainty is based on $\overline{e_c}$ (see above for the definition). We assume that the uncertainty in $\mu$ is proportional to $\overline{e_c}$. Since the values of $\overline{e_c}$ lie between 0.04 and 0.11, we take $\overline{e_c}$, for simplicity, as the percentage uncertainty of $\mu$. All uncertainties were treated symmetrically when constructing the error bars. Even when accounting for the error bars, the observed trend remains unchanged, indicating that the uncertainties do not affect the overall trend.}

\begin{figure}
    \centering    \includegraphics[width=\linewidth]{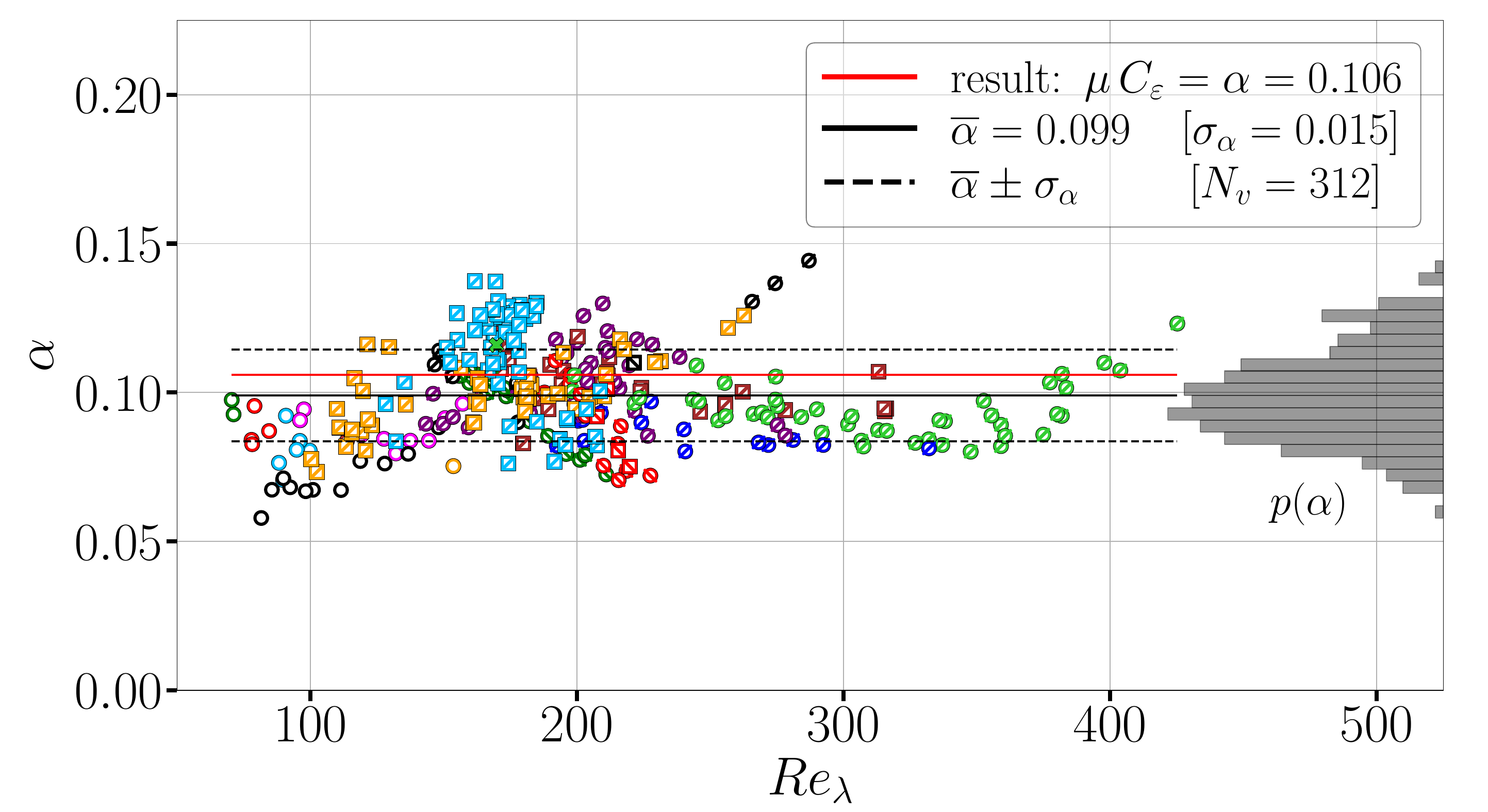} 
    \caption{\Vsix{$\alpha = \mu \, C_\varepsilon$ versus $Re_\lambda$ for the highly restricted data with the PDF of $\alpha$ values $p(\alpha)$. The red line indicates the result for $\alpha$ from the fit from figure~\ref{mu_c_epsi_e_error} while the black solid line represents the actual mean value of the ensemble of  $\alpha$ values. Additionally, two black dashed lines indicate the corresponding standard deviations $\sigma$ from the mean. The symbols and corresponding configurations are shown and explained in figure~\ref{mu_reynolds_num} and table~\ref{PhD measurements in LEGI 2023} in the appendix~\cite{SM}. $N_v$ indicates the number of velocity time series shown in this plot.}}
    \label{high_alpha_taylor_reynolds}
\end{figure}

\Vsix{In order to quantify the relation between $\mu$ and $C_\varepsilon$}, initially, a simple fit is presented through the data points. Note that the commonly accepted value of $\mu$ of around 0.26 is not the asymptotic limit of our measurements. Moreover, the product of $\mu$ and $C_\varepsilon$, here defined as $\alpha$, is shown as a function of $Re_\lambda$ in figure~\ref{high_alpha_taylor_reynolds}. Remarkably, $\alpha$ not only remains approximately constant for all cases, it also does not show any trends with $Re_\lambda$. Moreover, the well-centered PDF of $\alpha$ values indicates a Gaussian distribution. Table~\ref{alpha_values} shows that the value of the standard deviation of $\alpha$ remains lower than the individual corresponding values of the standard deviation of $C_\varepsilon$ and $\mu$. No turbulence quantity was identified that reduces the scatter perpendicular to the fitted curve. To the authors best knowledge, this is the first time a clear relation between these \Vsix{parameters} is found. 

\begin{figure}
    \centering    \includegraphics[width=\linewidth]{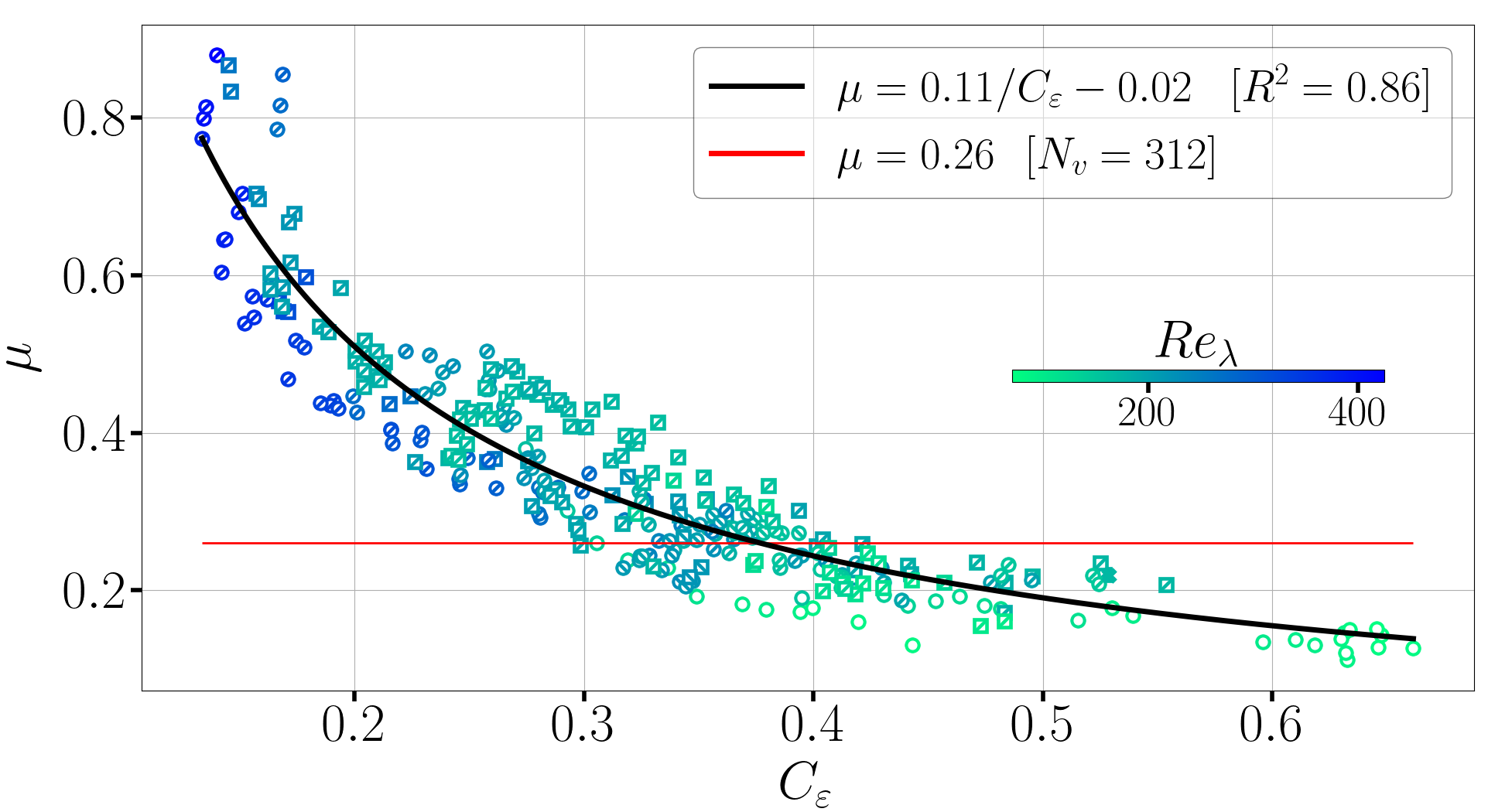} 
    \caption{\Vsix{$\mu$ as a function of $C_\varepsilon$ for the highly restricted dataset, where the red line indicates a commonly accepted value for $\mu$ for homogeneous isotropic turbulence~\cite{arneodo1996structure} and the black line corresponds to a least-square fit. The color indicates $Re_\lambda$. The symbols and corresponding configurations are shown and explained in figure~\ref{mu_reynolds_num} and table~\ref{PhD measurements in LEGI 2023} in the appendix~\cite{SM}. $N_v$ indicates the number of velocity time series shown in this plot.}}
    \label{c_epsi_e_mu_shape_factor_reynolds}
\end{figure}

\Vsix{Additionally, figure~\ref{c_epsi_e_mu_shape_factor_reynolds} presents $Re_\lambda$ for each velocity time signal as a color map over the relation between $\mu$ and $C_\varepsilon$ and shows that there is no clear dependence on $Re_\lambda$, as might be naively expected.} 

\begin{table} 
	\centering
	\begin{tabular}{llccccc}
		\toprule
		   restriction & fit result & mean ($\alpha$) & median ($\alpha$) & $\sigma_{\alpha}$ & $\sigma_{C_\varepsilon}$ & $\sigma_{\mu}$\\
		\midrule
		   High & 0.106 & 0.099 & 0.098 & 0.015 & 0.113 & 0.151 \\
                Low & 0.105 & 0.1 & 0.097 & 0.021 & 0.117 & 0.152 \\
		\bottomrule
	\end{tabular}
	\caption{\Vsix{For applying both the high and the low data selection criteria to the dataset, the fit results of the plots of $\mu$ over $C_\varepsilon$ as well as the mean, median and standard deviation $\sigma$ for the values of $\alpha$ are shown. In comparison, the standard deviation $\sigma$ of the ensemble values of $\mu$ over $C_\varepsilon$ are presented.}}
	\label{alpha_values}
\end{table}

\begin{figure}
    \centering    \includegraphics[width=\linewidth]{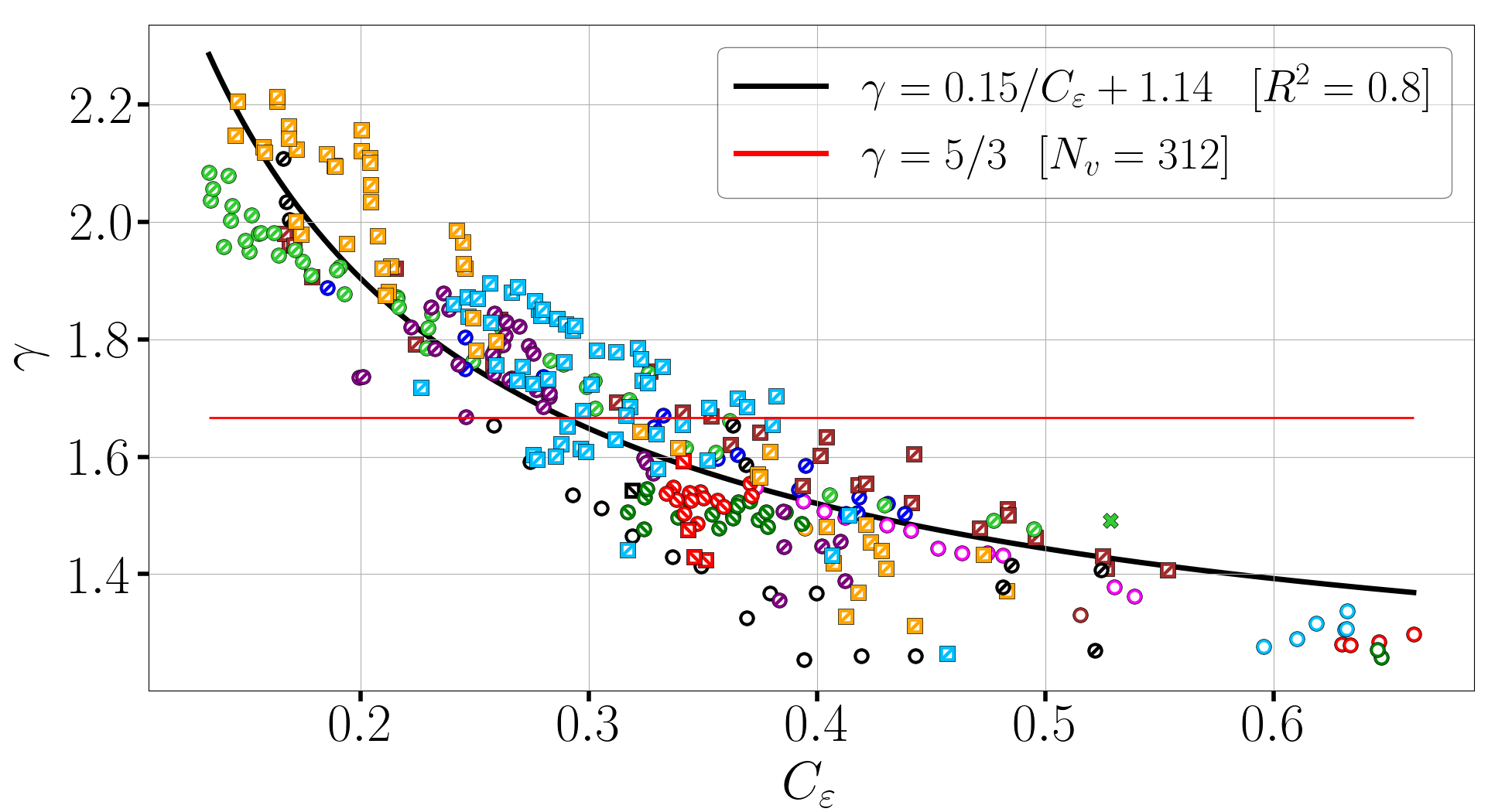} 
    \caption{\Vsix{$\gamma$ as a function of $C_\varepsilon$ for the highly restricted data. The red line indicates the commonly accepted value for $\gamma$ for HIT~\cite{sreenivasan1995universality}, while the black line corresponds to a least-square fit with $R^2$ being $0.8$. The symbols and corresponding configurations are shown and explained in figure~\ref{mu_reynolds_num} and table~\ref{PhD measurements in LEGI 2023} in the appendix~\cite{SM}. $N_v$ indicates the number of velocity time series shown in this plot.}}
    \label{gamma_c_epsi}
\end{figure}

\begin{figure}
    \centering    \includegraphics[width=\linewidth]{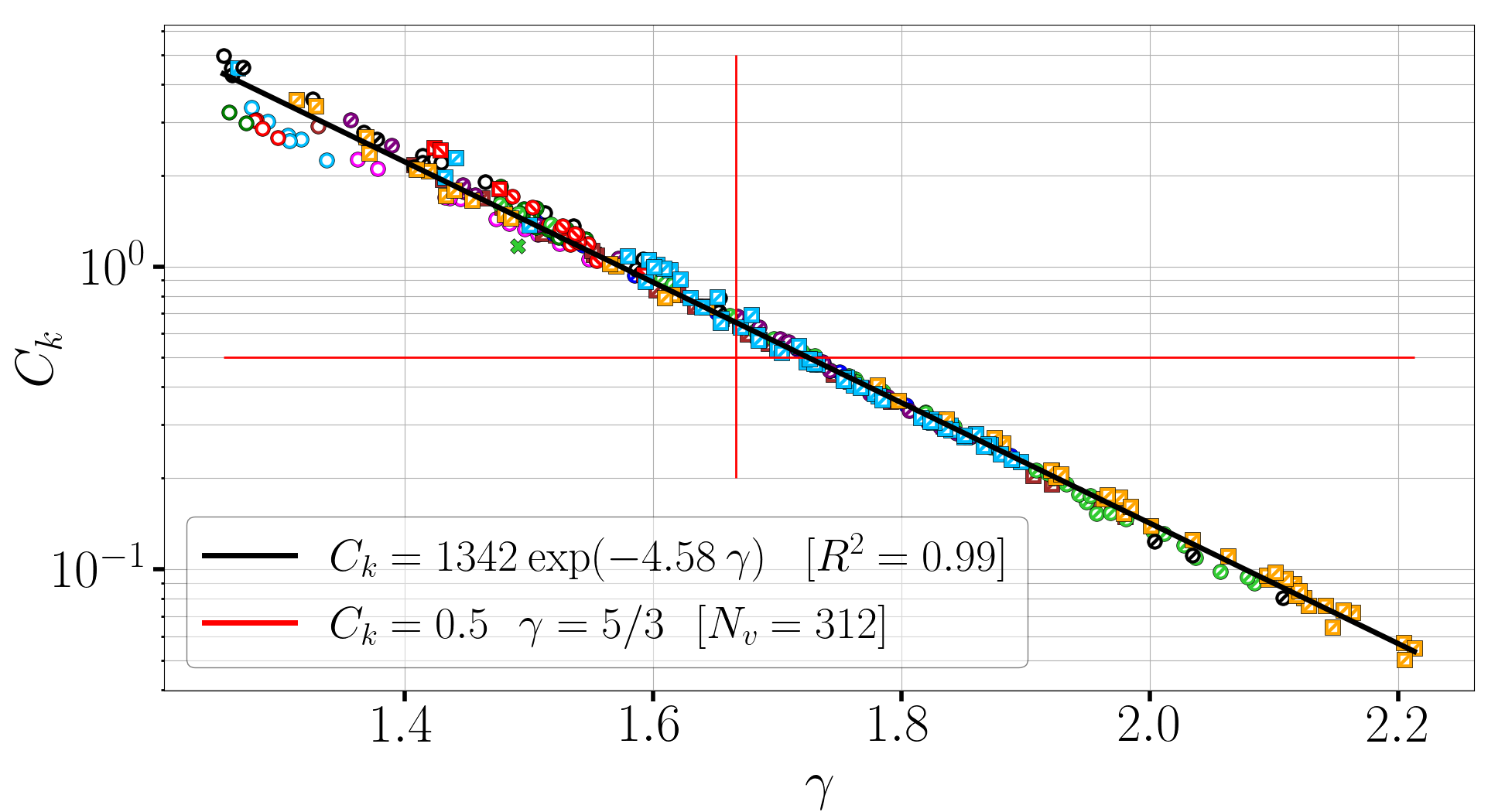} 
    \caption{\Vsix{$C_k$ versus $\gamma$ for the highly restricted data. The red lines indicate both the commonly accepted value for $C_k$ and $\gamma$ for HIT~\cite{sreenivasan1995universality}, while the black lines corresponds to a least-square fit with $R^2$ being $0.99$. Once more, the symbols and corresponding configurations are shown and explained in figure~\ref{mu_reynolds_num} and table~\ref{PhD measurements in LEGI 2023} in the appendix~\cite{SM}. $N_v$ indicates the number of velocity time series shown in this plot.}}
    \label{gamma_ck}
\end{figure}

\Vsix{We continue by investigating how $\gamma$ and $C_k$ behave within our dataset.} In figures~\ref{gamma_c_epsi} and \ref{gamma_ck}, the dependency of $\gamma$ on $C_\varepsilon$ as well as $C_k$ on $\gamma$ are presented, respectively ($C_k$ as a function of $C_\varepsilon$ is shown in the appendix~\cite{SM} in figure~\ref{ck_c_epsi}). Interestingly, the relation between $\mu$ and $C_\varepsilon$ is transferred to the other quantities. Moreover, all quantities present large variations with respect to the expected values for HIT within the K41 phenomenology. Since the present study focuses on the relationship between $\mu$ and $C_\varepsilon$, no further analysis is devoted to the dependencies between $\gamma$ and $C_\varepsilon$, nor between $\gamma$ and $C_k$. We remark that our results are still consistent with those reported in other experiments in inhomogeneous flows, such as other turbulent wakes~\cite{neunaber2020distinct} or atmospheric turbulence~\cite{rodriguez2023not}. Interestingly, Mydlarski \& Warhaft~\cite{mydlarski1996onset} also reported a relation between $\gamma$ and $C_k$ for active-grid-generated turbulence (and thus, a statistically steady HIT flow), which is included in figure~\ref{C_k_gamma_myd} in the appendix~\cite{SM}.

\section{Discussion \label{Dis}}

Based on the study of a very large number of velocity time series, the joint and individual behavior of the dissipation \Vsix{parameter} and the intermittency \Vsix{parameter} were quantified, as well as the spectral law of turbulence. In contrast to the common approach of more restrictive data selection to obtain more precise results on turbulence, we adopted a less restrictive data selection method. This follows the idea of Vassilicos~\cite{vassilicos2015dissipation}, which allows for the investigation of states of non-equilibrium turbulence. \Vsix{Regarding figure~\ref{mu_c_epsi_e_error}, our analysis leads to the proposition of a new relation,}
\begin{equation}
    \mu \: C_\varepsilon = \alpha = \text{const.},
    \label{mu_cepsilon_first}
\end{equation} 

\noindent \Vsix{extending} the previous proposition that these quantities should be individually constant \Vsix{for a wide range of canonical flows}. 

Figure~\ref{high_alpha_taylor_reynolds} indicates that there is no Reynolds number dependence of the product $ \mu \: C_\varepsilon$. This is a remarkable result, which is additionally supported by figure~\ref{c_epsi_e_mu_shape_factor_reynolds} \Vsix{where the value of $Re_\lambda$ is not unique for neither $\mu$ nor $C_\varepsilon$. This becomes even clearer when looking at figure~\ref{c_epsi_e_mu_shape_factor_all_reynolds} in the appendix~\cite{SM}, where the plotted data is extended by only applying low restrictions. Surprisingly, eq.~(\ref{mu_cepsilon_first}) also holds for the low-restricted dataset, although the scatter is increased, as shown in figures~\ref{c_epsi_e_mu_shape_factor_all} and \ref{low_alpha_taylor_reynolds} displayed in the appendix~\cite{SM}. Moreover, we provide evidence in figure~\ref{castaing_mu_c_epsi_e} in the appendix~\cite{SM} that the relation between $\mu$ and $C_\varepsilon$ is independent of the Castaing error.}

\Vsix{As noted above, $\alpha$ is not only independent of $Re_\lambda$ but we also did not identify any parameter that reduces its global scatter. This indicates that the relation from eq.~(\ref{mu_cepsilon_first}) is independent of any additional quantity for stationary 1$\,$D turbulence under the given selection criteria and assumptions. This is further strengthened by the fact that the distribution of $\alpha$ is close to Gaussian, indicating that the dispersion in $\alpha$ occurs from random uncertainties. Furthermore, table~\ref{alpha_values} reveals that the spreading around the mean for $\alpha$ is approximately one order of magnitude lower than for both $C_\varepsilon$ and $\mu$.}

It has long been claimed that certain turbulent properties are universal, $e.g.$, independent of the type of flow, for large enough Reynolds numbers. Also, for free-shear anisotropic flows, a ``tendency of recovery" of local isotropy is expected at high Reynolds numbers ($cf.$ \cite{zhou2021turbulence}). Together with the relations to other turbulence \Vsix{parameters} (shown in figures~\ref{gamma_c_epsi} and \ref{gamma_ck}), this new proposed relation combines the dimensionless energy dissipation as an overall quantity of turbulent states with the statistics of velocity fluctuations at two points. Note that the power spectrum and the intermittency \Vsix{parameter} fully capture the two-point statistics~\cite{morales2012characterization}. Thus, we observe a new universality combining overall \Vsix{parameters} ($C_\varepsilon$ or $C_k$) with the two-point statistics of the velocity field. \Vsix{In the appendix~\cite{SM}, figures~\ref{gamma_c_epsi_all} and \ref{C_k_gamma_myd} show that the relations in figures~\ref{gamma_c_epsi} and \ref{gamma_ck} also hold for applying less restrictions.}

As a remark, it should be pointed out that our results do not imply the absence of Reynolds number effects. For example, $L/\eta$ is still expected to depend on $Re_\lambda$, influencing the range of the power law of the power spectrum and therefore the \Vsix{non-Gaussianity} at the smallest scales (which are also expected to be determined by the Reynolds number). We also observe that the evolution of $C_\varepsilon$ with the Reynolds number follows different linear relations sections for different flows, as discussed above. \Vsix{In this context, it is instructive to examine the Reynolds-number dependence of the individual parameters in more detail. Barenblatt \& Goldenfeld~\cite{barenblatt1995does} argued that even for fully developed turbulence, $C_2$ as the prefactor of the law governing the inertial range of the second-order structure function, should scale as (ln$(Re_\lambda))^{-1}$. In contrast, our data suggests for one set of boundary conditions, that $C_\varepsilon$ approximately scales linearly with $1/Re_\lambda$, while both $\mu$ and $\gamma$ exhibit an approximately linear dependence on $Re_\lambda$. Moreover, the logarithm of $C_k$ approximately scales linearly with $Re_\lambda$. For verification, figure~\ref{ck_taylor_reynolds_num} in the appendix~\cite{SM} shows $C_k$ as a function of $Re_\lambda$.}

\Vsix{Furthermore, also the range of $\mu$-values warrants a brief discussion, particularly for values exceeding 0.5. To the best of our knowledge, previous studies have not reported values higher than 0.5. However, this is not unexpected, as most existing work focuses on HIT. Furthermore, as there is no theoretical derivation of $\mu$, there are also no
upper limits known. Figure~\ref{object_distance} shows that values of $\mu > 0.5$ occur exclusively for cylinder wakes at streamwise distances smaller than 20$~d$ within our dataset. This suggests that elevated values of $\mu$ may be associated with increased shear.}

\Vsix{The estimation of $\mu$ requires a pronounced inertial range. The extent of the inertial range can be estimated by the ratio of the integral length scale $L$ and the Kolmogorov length scale $\eta$. The relation between $\mu$, $Re_\lambda$ and the ratio $L/\eta$ is presented in figure~\ref{mu_L_div_eta}. It can be seen that the value of $\mu$ is not influenced by the value of $L/\eta$.
As a remark, it should be pointed out that the range of $L/\eta$ presented in the work is similar to the one reported by Mydlarski \& Warhaft~\cite{mydlarski1996onset}. Note also that, under the definition of the integral length scale used here, where the integration of the autocorrelation function is made until $1/e$, $L$ is proportional to the value obtained with more standard methods, but systematically smaller. In consequence, values of $L/\eta$ reported here are underestimated by at least a factor 2 when compared with other standard methods.}

\Vsix{As mentioned, Mydlarski \& Warhaft also estimated $C_\varepsilon$, $\mu$, $\gamma$ and $C_k$ in 1996~\cite{mydlarski1996onset}}. For $\mu$, $\gamma$ and $C_k$, they reported Reynolds-number dependencies on $Re_\lambda$ that are similar to those found in our work, for example, both $\mu$ and $\gamma$ increase with $Re_\lambda$. However, their dataset was limited in both the number of time series and the number of grid-generated experimental configurations (mostly using an active grid and thus close to HIT). For instance, their values of $C_\varepsilon$ are reported to be constant, in contrast to our findings. Since all four quantities are computed somewhat differently, and we additionally filter for non-Gaussianity at large scales, the direct comparability of both studies remains unclear. In contrast to the present work, Mydlarski \& Warhaft aimed to determine the behavior of $\gamma$ and $C_k$ as $Re_\lambda \rightarrow \infty$ within HIT. Overall, our work can be seen as an extension of the study of Mydlarski and Warhaft to inhomogeneous turbulence~\cite{mydlarski1996onset}. Later, Cleve $et \, al.$ reported a similar trend of $\mu$ on $Re_\lambda$ in comparison to our work, using data from atmospheric measurements and shear flows~\cite{cleve2004intermittency}.

Next, we set our result in the context of the $\Pi$-theorem of Vaschy-Buckingham~\cite{vaschy1892lois, buckingham1914physically}. $C_\varepsilon$ and $\mu$ can also be viewed as two $\Pi$-parameters for dissipation and the scale dependency of the fluctuations of dissipation. For different turbulent flow configurations, it is a challenge to find the functional relation between these $\Pi$-parameters. For a prominent example, see the Rayleigh-B{\'e}nard convection~\cite{ahlers2009heat}. In this sense, our result reads as $ \mu \: C_\varepsilon = \Pi_1 \cdot \Pi_2 = \text{const.} \neq \mathrm{func}(Re)$. 

Further, a relevant consequence of eq.~(\ref{mu_cepsilon_first}) is that we can define $\mu$ as,
\begin{equation}
    \mu = \frac {u^{\prime 3} \: \alpha} {\varepsilon \: L}.
    \label{mu_cepsilon_second}
\end{equation}
This relation has the advantage that the quantities on the right-hand side are much easier to measure and converge than $\mu$ (when estimated using standard approaches). 

Next, we take a closer look at the quantities $\mu$ and $C_\varepsilon$ to discuss their meaning in the cascade process. By re-writing eq.~(\ref{mu}) as,

\begin{equation}
    \mu \propto \frac{\mathrm{d \:\:}\Lambda^2(r)}{\mathrm{d \:\: ln}(r)},
    \label{mu_derivation}
\end{equation}

\noindent the intermittency \Vsix{parameter} $\mu$ can be interpreted as the “speed” of the cascade — that is, how rapidly (starting at the largest scales and then going to smaller scales) the shape of the normalized velocity increment PDF evolves from Gaussian behavior at large scales to heavy-tailed distributions at small scales. The concept of cascade speed \Vsix{as an analogy} is not new and has been used previously in the literature (see for instance the work of Siefert \& Peinke~\cite{siefert2004different}) and is further explained in the \Vsix{following}.

\begin{figure}
    \centering    \includegraphics[width=\linewidth]{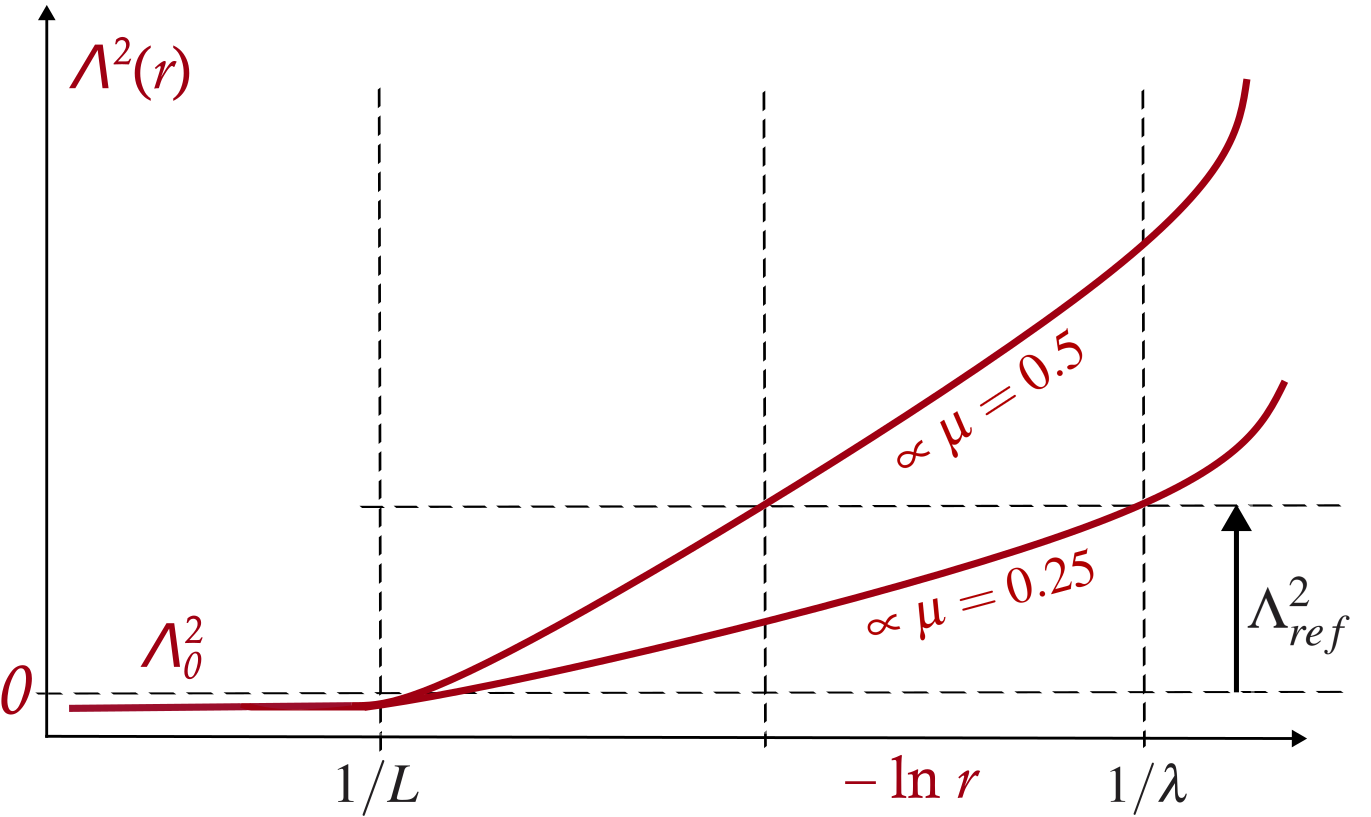} 
    \caption{\Vsix{Schematic representation of the spatial expansion of the turbulence cascade: the red curves illustrate two different cases of the evolution of the shape \Vsix{parameter} $\Lambda^2(r)$ for normalized velocity increments. While one curve exhibits the behavior of a moderate value of $\mu = 0.5$, the other curve shows the behavior of a higher value of $\mu = 0.25$.  $L$ is the integral length scale, $\lambda$ is the Taylor length scale and $\Lambda^2_{ref}$ indicates a reference value for $\Lambda^2$. The Note that in this sketch, $\Lambda_0^2$ is approximately zero.}}
    \label{speed_cascade}
\end{figure}

\Vsix{Figure~\ref{speed_cascade} illustrates different behaviors of the shape parameter corresponding to distinct $\mu$ values, based on figure~\ref{Turbulence_scheme}. We employ the terminology of cascade speed strictly in the sense of the evolution of the shape \Vsix{parameter}. Note that this speed does not represent a physical velocity (measured in meters per second), but rather serves as an analogy for the rate at which the cascade process proceeds. For simplicity, these cases in figure~\ref{speed_cascade} share identical values of $L$ and $\lambda$. Starting from eq.~(\ref{mu}), which reflects experimentally observed behavior, and applying the selection criteria $\Lambda^2_0 \approx 0$ leads to}

\begin{equation}
    \Vsix{
    {\Lambda^2(r) = \mu/9 \; \ln(L/r).}
    \label{speed_1}
    }
\end{equation}

\noindent \Vsix{For the purpose of illustration, we additionally define $\Lambda^2_{ref}$, a shared target value for $\Lambda^2$ (indicated in figure~\ref{speed_cascade}). Eq.~(\ref{speed_1}) can therefore be modified to}

\begin{equation}
    \Vsix{
   {\frac{\Lambda^2_{ref}}{\mu} \propto \ln(L/r).}
    \label{speed_2}
    }
\end{equation}

\noindent \Vsix{As $L$ and $\Lambda^2_{ref}$ are identical for both curves, eq.~(\ref{speed_2}) can therefore be simplified to}

\begin{equation}
    \Vsix{
    {\frac{1}{\mu} \propto \ln(1/r).}
    \label{speed_3}
    }
\end{equation}

\noindent \Vsix{Initiating the cascade at the scale $L$, the upper curve corresponding to $\mu = 0.5$ reaches $\Lambda^2_{ref}$ at a comparatively high value of $r$ and is thus considered faster, or respectively with higher ``speed". In contrast, the lower curve corresponding to $\mu = 0.25$ reaches $\Lambda^2_{ref}$ at a comparatively low value of $r$ and is thus considered slower. In other words, the cascade is considered faster when a given value of $\Lambda^2$ is reached over a smaller range of scales and $vice \: versa$.}

Regarding $C_\varepsilon$ and following Apostolidis $et \, al.$~\cite{apostolidis2022scalings}, the dissipation parameter can be re-expressed (introducing a large scale energy density rate $\varepsilon_L$) as,

\begin{equation}
    C_\varepsilon  = \frac{\varepsilon}{\varepsilon_L} = \frac{\varepsilon}{\varepsilon_L (\varepsilon)},
    \label{cepsilon_efficiency}
\end{equation}

\noindent by combining $L$ and $u^{\prime 3}$, leading to $\varepsilon_L = u^{\prime 3}/L$. In this formulation, $C_\varepsilon$ represents the ratio between the actual dissipation rate $\varepsilon$ and $\varepsilon_L$, that acts as a surrogate of the input energy flux, associated with large-scale forcing. $C_\varepsilon$ therefore quantifies the energy ratio of the turbulence cascade process converting large-scale input into small-scale dissipation. 

\begin{figure}[h!] 
    \centering
    \includegraphics[width=\linewidth]{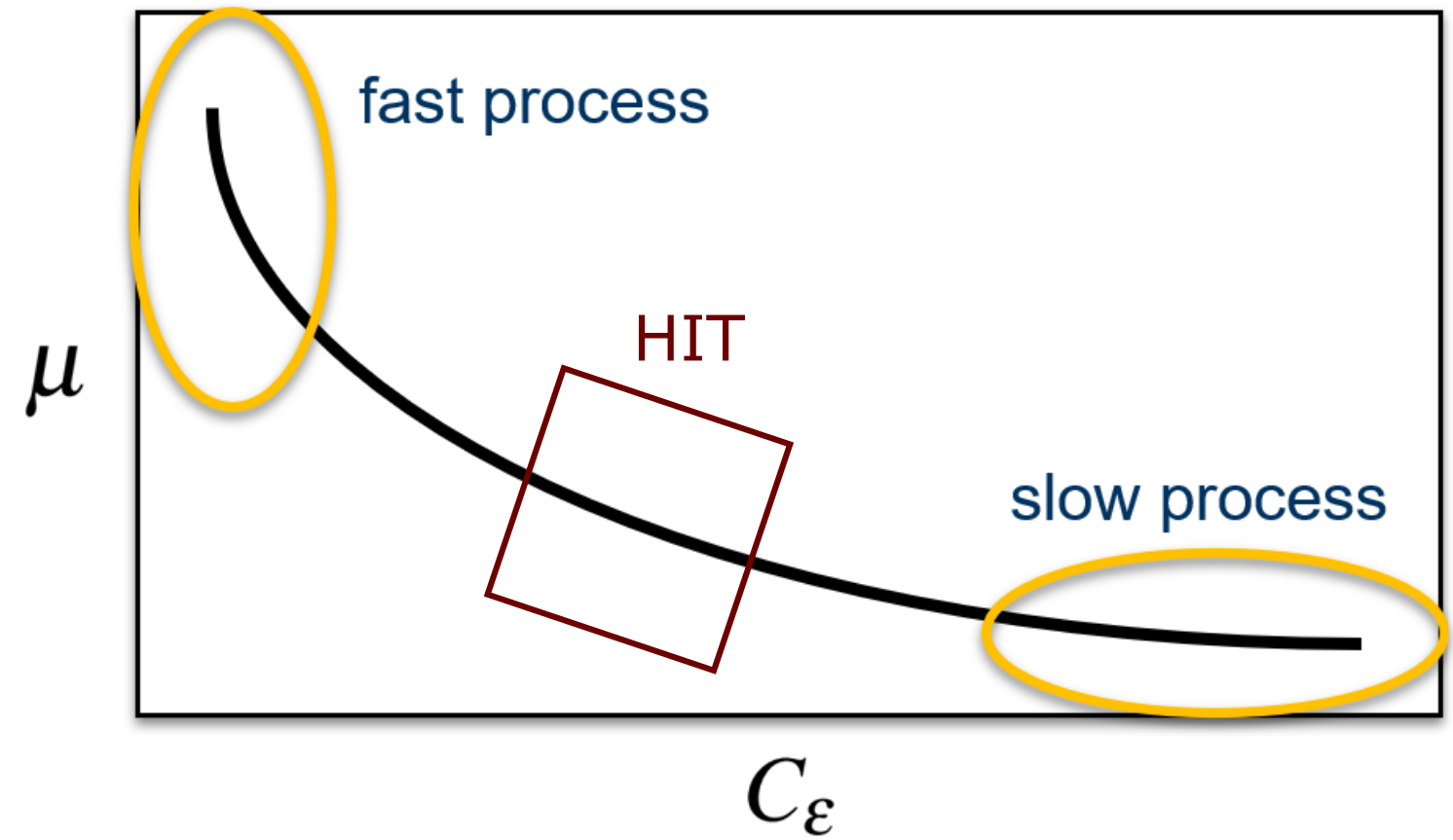} 
    \caption{Schematic representation of the relation between $\mu$ and $C_\varepsilon$. The yellow ellipses highlight the two boundary regions in the spectrum of $\mu$ and $C_\varepsilon$ values, whereas the dark red square indicates the region of HIT.}
    \label{Turbulence_scheme_quantities}
\end{figure}

Given these interpretations, the observed relationship between $\mu$ and $C_\varepsilon$ implies that a faster cascade tends to have a lower energy ratio of the turbulence cascade process converting large-scale input into small-scale dissipation. Turbulence, being a fundamentally non-equilibrium process, can be characterized thermodynamically by how far it deviates from equilibrium. We refer here to the thermodynamic equilibrium in the strict sense, $e.g.$ the state of maximum entropy with no macroscopic fluxes, not to be confused with the equilibrium definition proposed in the context of ``non-equilibrium turbulence'', which is discussed in the introduction. In this context, the values of $C_\varepsilon$ (and $\mu$) serve as a measure of this deviation. Figures~\ref{Turbulence_thermo} and \ref{Turbulence_scheme_quantities} illustrate this behavior: as the system moves further from thermodynamic equilibrium, the cascade becomes faster (larger $\mu$), dissipation increases (larger $\varepsilon$), but the energy ratio decreases (smaller $C_\varepsilon$), and vice versa. Note that, unlike $C_\varepsilon$ and $\mu$ the variation of $\varepsilon$ considered here refers to comparisons to other measurement points within the same set of boundary conditions. \Vsix{Moreover, it should be noted that $\varepsilon$ is not proportional to $C_\varepsilon$, as $\varepsilon_L$ in eq.~(\ref{cepsilon_efficiency}) depends on $\varepsilon$. In fact, due to this dependence, $\varepsilon$ varies inversely with $C_\varepsilon$ (as indicated in figure~\ref{Turbulence_thermo}}). Nevertheless, while the reported values of both $\mu$ and $C_\varepsilon$ concern values with both identical and different boundary conditions, we find a collapse that expands across all studied cases. 

Overall, the coupling between $\mu$ and $C_\varepsilon$ appears to be intrinsic to turbulence and opens a new perspective on the field, suggesting that turbulence both within and beyond HIT is governed by the same cascade mechanism. However, a physical origin of the relationship between $\mu$ and $C_\varepsilon$ and also between $\gamma$ and $C_k$ remains an open question. \Vsix{We are not able to derive the results from first principles, such as the the Navier--Stokes equations or the laws of thermodynamics. For now, these results remain purely empirical.} 

\Vsix{As noted above, a preliminary analysis indicates that our results remain valid even under stronger shear, provided that the large-scales statistics are approximately Gaussian. We reserve a comprehensive analysis of the full dataset, comprising more than 5000 velocity time series, for a follow-up study in which all parameters will be systematically assessed for consistency. Nevertheless, further work is required to establish the general validity of these findings in turbulence. In particular, it remains to be tested whether they extend to other flows, such as wall-bounded flows or atmospheric turbulence. The analysis should also be extended to higher-dimensional measurements, also taking transverse increments into account. In particular under quantification of shear and vorticity, for which we expect more directional dependencies. In this framework, the turbulent kinetic energy may also be employed instead of $u^\prime$, as recently suggested~\cite{vassilicos2015dissipation}. In addition, temporal aspects should be incorporated, as stationarity has been argued to play a role in the emergence of intermittency\cite{aumaitre2024intermittency}.}

\Vsix{Despite the lack of a theoretical foundation, our results are robust across a large variety of flow configurations, indicating that large scales ($L, u^\prime$), inertial range dynamics ($\mu$), and dissipation scales ($C_\varepsilon$) exhibit coupling and seem to adapt to the flow conditions.}

\begin{figure}[h]
\centering
\newcommand{\block}[2]{%
    \begin{array}{c}
        \text{#1} \\#2
    \end{array}}
\[
\begin{array}{c}

\block{fast process}{\mu \uparrow}
\quad \widehat{=}\quad
\block{high dissipation}{\varepsilon \: (\mathrm{case \; specific}) \uparrow}
\quad \widehat{=}\quad
\block{low energy ratio}{ C_\varepsilon \downarrow}
\\
\noalign{\vskip 6pt}
\hline
\noalign{\vskip 6pt}

\block{slow process}{\mu \downarrow}
\quad \widehat{=}\quad
\block{low dissipation}{\varepsilon \: (\mathrm{case \; specific}) \downarrow}
\quad \widehat{=}\quad
\block{high energy ratio}{ C_\varepsilon \uparrow}
\end{array}
\]
\caption{Relations of the intermittency \Vsix{parameter} $\mu$, the mean dissipation rate $\varepsilon$ and the dissipation \Vsix{parameter} $C_\varepsilon$.}
    \label{Turbulence_thermo}
\end{figure}

\clearpage

\appendix

\section{Further analysis of the dataset \label{SM}}

In this appendix, we complement the main text showing a detailed example of our data analysis protocol. Additionally, further relations between \Vsix{parameters} are also shown and discussed, including a discussion about the validity of our results for our dataset processed with reduced quality constraints (in the sense of the restrictions discussed in section~\ref{Res}).

\section*{Castaing curve} 

\Vsix{The Castaing curves can be constructed by applying eqs.~(\ref{castaing_curve}) and~(\ref{castaing_curve_sigma})}

\begin{equation}
\begin{aligned}
{\hat{p}(u_r(x)/\sigma_{r})} = \\
\frac{1}{2 \pi \Lambda(r)} \:
\int_{0}^{\infty} \: 
\frac{\mathrm{d}\sigma}{{\sigma}^2} \:
\exp \left[- \frac{{u_r(x)}^2}{{2\sigma}^2} \right] \:
\exp \left[ -\frac{\mathrm{\ln}^2 \left({\sigma/\sigma_0}\right)}{2\Lambda^2(r)} \right]
\end{aligned}
\label{castaing_curve}
\end{equation}

\begin{equation}
\sigma_{0}^2 = \overline{u_r(x)^2} \:
\exp \left[- 2\Lambda^2(r) \right]
\label{castaing_curve_sigma}
\end{equation}

\section*{Characterization of the experimental cases}
\begin{figure}[h!] \centering
    \includegraphics[width=\linewidth]{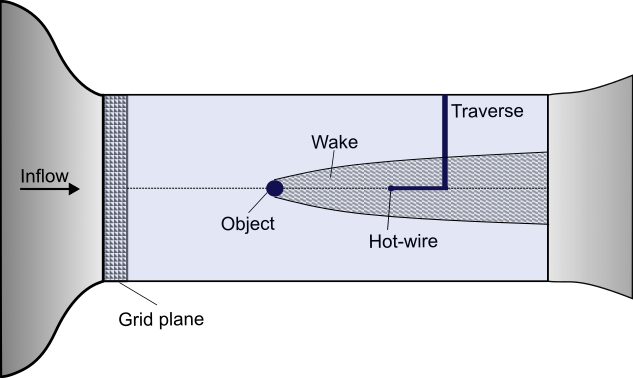} 
    \caption{Side view of the experimental setup in LEGI lab. The hot wire can be traversed in the plane from the figure both in streamwise and spanwise directions (only centerline data is presented in this work). Both the grid and the object can be exchanged or removed individually.}
    \label{Wind_tunnel_Grenoble}
\end{figure}

\Vsix{Figure~\ref{Wind_tunnel_Grenoble} shows the general experimental setup in LEGI. Table~\ref{PhD measurements in LEGI 2023} presents an overview of all cases conducted in the LEGI wind tunnel in 2023.}

\begin{table*} 
	\centering
	\begin{tabular}{lcccccccccc}
		\toprule
		   & $\textrm{L}_1$ & $\textrm{W}_1$ & $\textrm{W}_2$ & $\textrm{T}_2$& $\textrm{\cite{ferran2023characterising}}$ & $\textrm{\cite{mora2019energy}}$ & $\textrm{\cite{renner2001experimental}}$\\
		\midrule
		Cylinder, $ \o  =$ 10\,mm, $s=100\,$\%, $Re \approx 6\,$k & C1 & - & - & -& - & - & -\\
		Cylinder, $\o  =$ 20\,mm, $s=100\,$\%, $Re \approx 13\,$k & C2 & C3 & C4 & -& - & - & -\\
		Cylinder, $\o  =$ 20\,mm, $s=58\,$\%, $Re \approx 13\,$k  & C5 & - & C6 & -& - & - & -\\
		Cylinder, $\o  =$ 20\,mm, $s=100\,$\%, $Re \approx 8\,$k  & - & - & C7 & C8 & - & - & - \\
		Cylinder, $\o  =$ 20\,mm, $s=100\,$\%, $Re \approx 5\,$k  & - & - & - & C9 & - & - & -\\
		
		Disk, $\o  =$ 72\,mm, $s=34\,$\%, $Re \approx 47\,$k  & D10 & - & D11 & -& - & - & -\\
		Disk, $\o  =$ 72\,mm, $s=48\,$\%, $Re \approx 47\,$k & D12 & - & D13 & -& - & - & -\\
		Disk, $\o  =$ 72\,mm, $s=34\,$\%, $Re \approx 28\,$k & D14 & -  & -  & -& - & - & -\\
		Disk, $\o =$  72\,mm, $s=48\,$\%, $Re \approx 28\,$k  & -  & - & -  & D15 & - & - & -\\
		
		Grid, $\overline{u}_\infty=10\,$m/s  & - & G16 & G17 & -& - & - & -\\
		Grid, $\overline{u}_\infty=6\,$m/s  & - & - & G18 & G19 & - & - & -\\

        Grid, $\overline{u}_\infty=7\,$m/s  & - & - & -& -& G20 & - & -\\
		Grid, $\overline{u}_\infty=5\,$m/s  & - & - & -& -& G21 & - &\\
        Grid, $\overline{u}_\infty=4\,$m/s  & - & - & -& -& G22 & - & -\\
		Grid, $\overline{u}_\infty=8.6\,$m/s  & - & - & -& -& - & G23 & - \\
        Grid, $\overline{u}_\infty=17\,$m/s  & - & - & -& -& - & G24 & - \\
        Jet, $\overline{u}_\infty=45.5\,$m/s  & - & - & -& -& - & - & J25 \\
		
		\bottomrule
	\end{tabular}
	\caption{All measurement configurations used in this study. The first eleven lines correspond to the turbulent wake measurements. Within those, the first nine lines correspond directly to wake measurements and the two following lines correspond to the background flow measurements in the wind tunnel. The remaining lines of the table correspond to the studies based on other flows extracted from previous works. The column descriptions are sorted as follows: ``L'' means that there is no grid, ``W'' stands for a static grid which is made out of cylindrical wooden bars with a diameter of $20\,$mm, ``T'' stands for triple random mode of the active grid where all the bars are driven independently with independent rotational speed and in independent direction. The indices 1 and 2 are standing for the distance between the begin of the test section (= grid plane in figure~\ref{Wind_tunnel_Grenoble}) and the object ($1 = 630\,$mm, $2 = 1610\,$mm). The last three columns correspond to the studies from the literature~\cite{ferran2023characterising, mora2019energy, renner2001experimental}, respectively. For the actual wake measurements, $\o$ is the diameter of the wake-generating object, $s$ is the solidity of the object and $Re$ is the Reynolds number depending on the object diameter and the inflow velocity. For all other measurements, $\overline{u}_\infty$ is the mean inflow velocity.}
	\label{PhD measurements in LEGI 2023}
\end{table*}

\section*{Example velocity time series} 

\Vsix{Table~\ref{data_example} shows the characteristics of the velocity time series used as an example in the main manuscript}

\begin{table*} 
	\centering
        \setlength{\tabcolsep}{2pt} 
        \scriptsize
	\begin{tabular}{lccccccccccccccccccc}
		\toprule
		   $T$ (s) & $f$ (kHz) &$x/d$ & $\overline{u}$ (m/s) & $u^{\prime}/\overline{u}$ ($\%$) & $Re_{\lambda}$ & $S_u$ & $F_u$ & $L$ (cm)& $\lambda$ (mm)& $\eta$ (mm) & $\varepsilon$ (m²/s³) & $C_\varepsilon$ & $\mu$ & $\alpha$ & ${\Lambda}^2_0$ & $\gamma$ & $C_k$ \\
		\midrule
		   120 & 50 & 5 & 6.55 & 22.26 & 533 & 0.23 & 3.11 & 7.75 & 5.6 & 0.12 & 15.6 & 0.39 & 0.27 & 0.106 & 0.005 & 1.59 & 0.88\\
		\bottomrule
	\end{tabular}
	\caption{The characteristics of the used velocity time signal in figure~\ref{scheme_data_example} are shown: $T$ is the sampling time, $f$ is the sampling frequency, $x/d$ is the streamwise distance normalized by the disk diameter, $\overline{u}$ is the mean velocity at the acquisition point, $u^{\prime}/\overline{u}$ is the turbulence intensity, $Re_{\lambda}$ is the Reynolds number based on the Taylor length scale, $S_u$ is the skewness of the velocity time signal, $F_u$ is the flatness of the velocity time signal, $L$ is the integral length scale, $\lambda$ is the Taylor length scale, $\eta$ is the Kolmogorov length scale, $\varepsilon$ is the mean dissipation rate, $C_\varepsilon$ is the dissipation \Vsix{parameter}, $\mu$ is the intermittency \Vsix{parameter}, $\alpha$ is the product of $\mu$ and $C_\varepsilon$, ${\Lambda_0}^2$ is the shape \Vsix{parameter} on the very large scales, $\gamma$ is the negative slope of the energy spectrum within the inertial range and $C_k$ is the Kolmogorov \Vsix{parameter} (according to eq.~(\ref{energy_spectrum})).}
	\label{data_example}
\end{table*}

\section*{Relations between non-dimensional quantities with less restrictions}

\begin{figure}[h!] 
    \centering    \includegraphics[width=\linewidth]{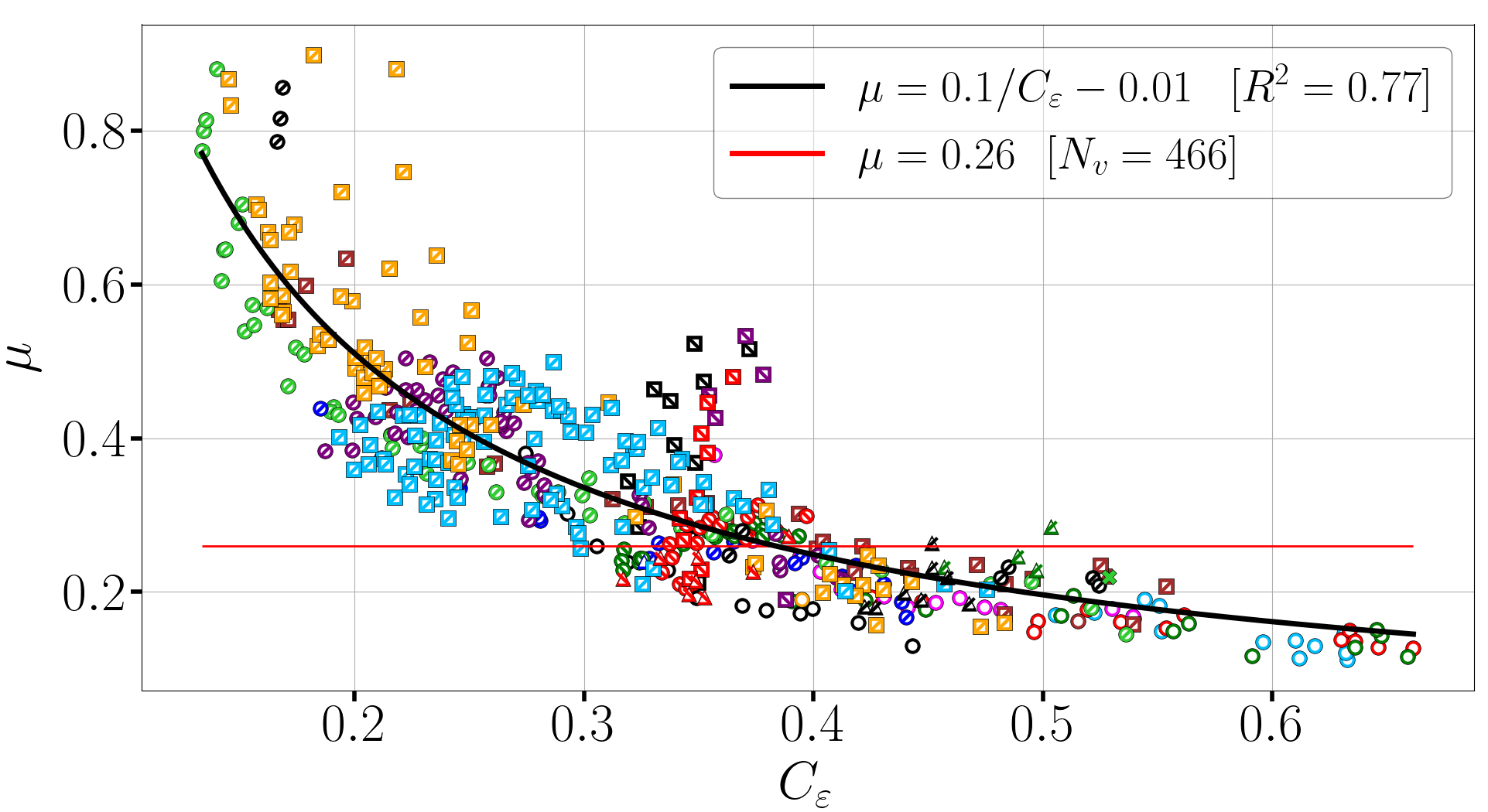} 
    \caption{$\mu$ over $C_\varepsilon$ for the less restricted dataset. The red line indicates a commonly accepted value for $\mu$ for homogeneous isotropic turbulence~\cite{arneodo1996structure}, while the black line corresponds to a least-square fit ($R^2= 0.77$). The symbols and corresponding configurations are shown in figure~\ref{mu_reynolds_num} from the main manuscript and explained in table~\ref{PhD measurements in LEGI 2023}. $N_v$ indicates the number of velocity time series shown in this plot.}
    \label{c_epsi_e_mu_shape_factor_all}
\end{figure}

\begin{figure}
    \centering    \includegraphics[width=\linewidth]{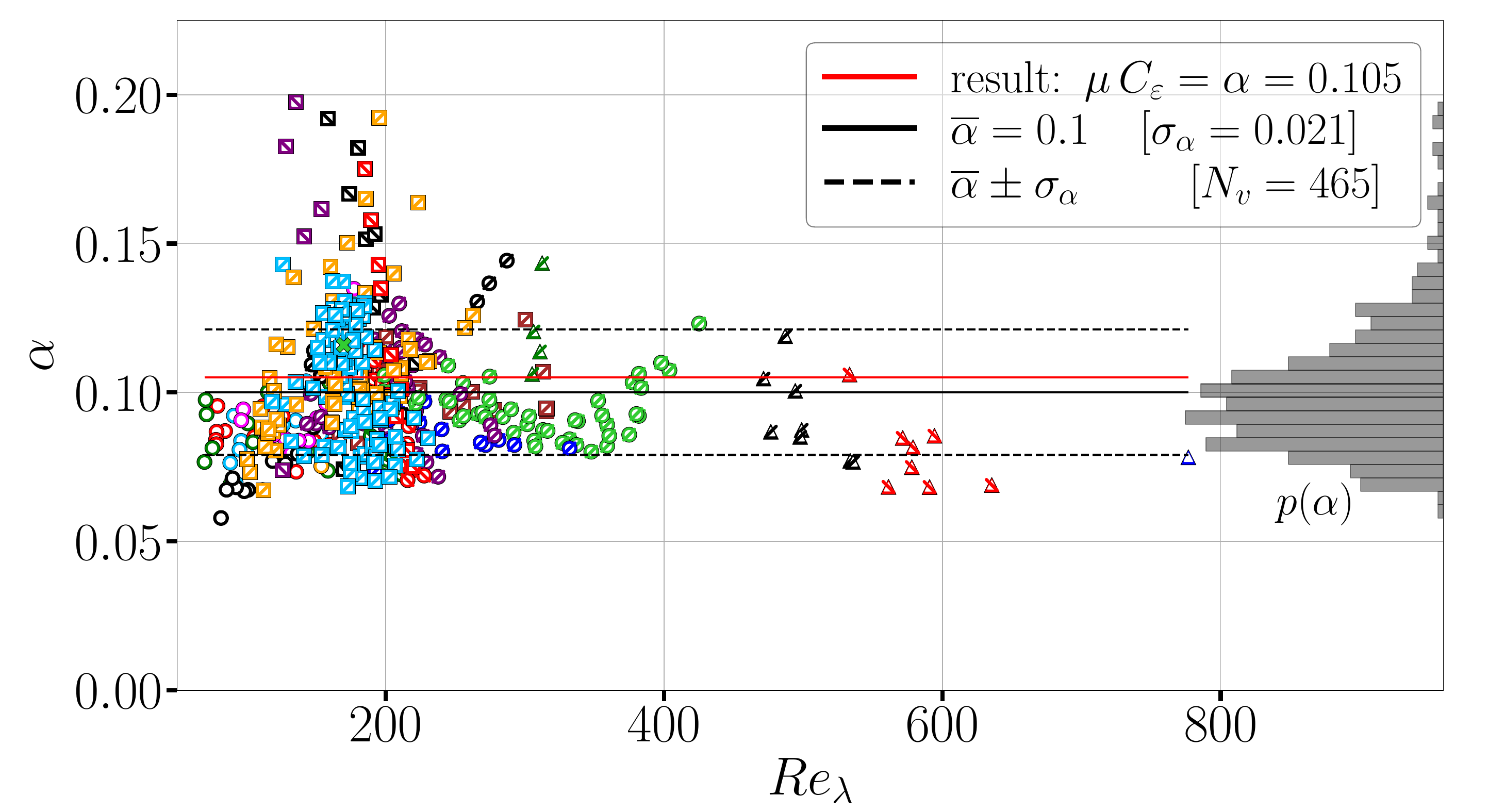} 
    \caption{For all data points within the less restricted dataset, $\alpha$ is presented over $Re_\lambda$ with the PDF of $\alpha$ values $p(\alpha)$. The red line indicates the result for $\alpha$ from the fit in figure~\ref{c_epsi_e_mu_shape_factor_all}. The black solid line represents the actual mean value of all plotted $\alpha$ values. Additionally, two black dashed lines indicate the corresponding standard deviations $\sigma$ from the mean. The symbols and corresponding configurations are shown in figure~\ref{mu_reynolds_num} from the main manuscript and explained in table~\ref{PhD measurements in LEGI 2023}. $N_v$ indicates the number of velocity time series shown in this plot.}
    \label{low_alpha_taylor_reynolds}
\end{figure}

\begin{figure}
    \centering    \includegraphics[width=\linewidth]{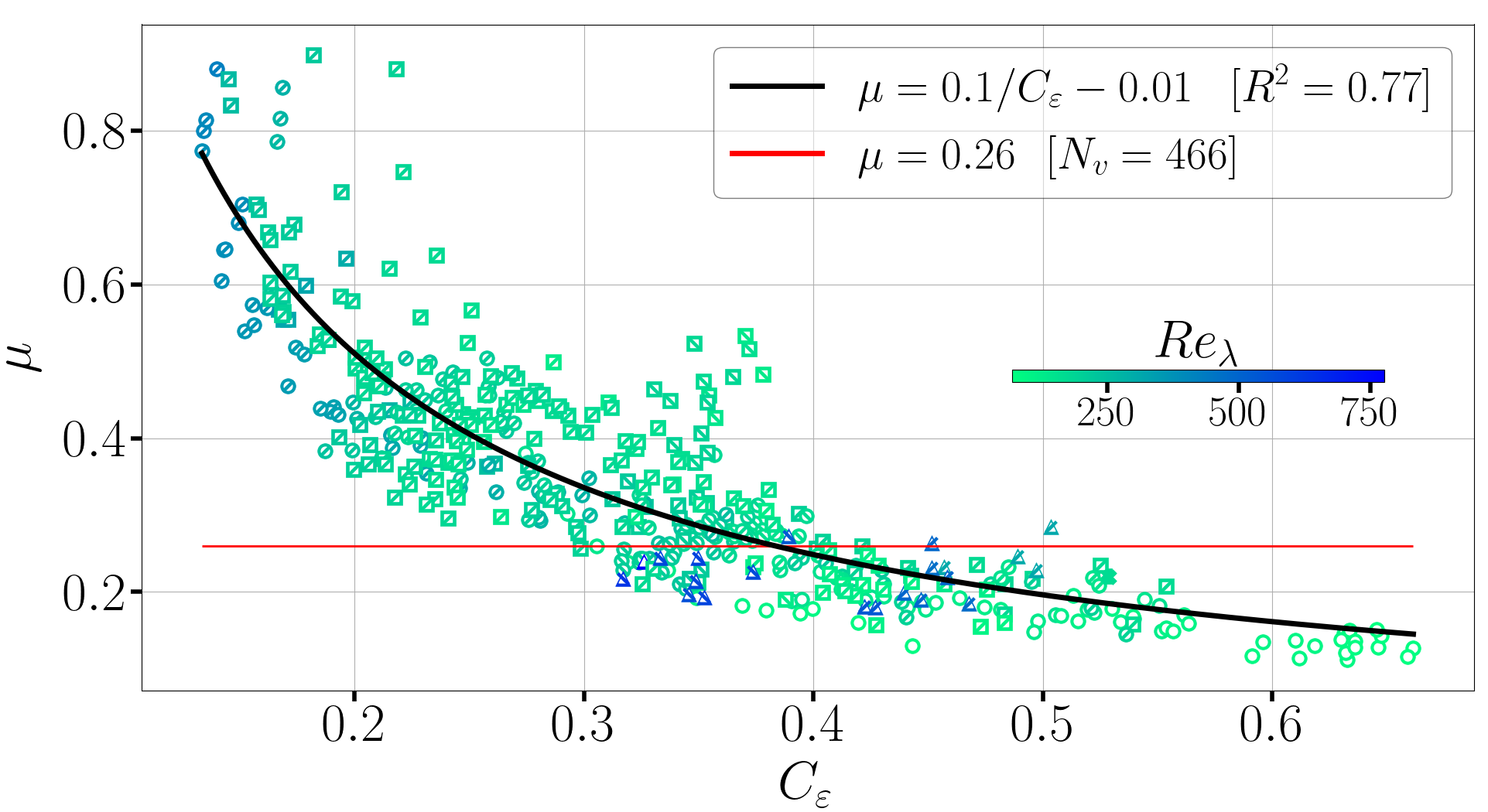} 
    \caption{For all data points within the less restricted dataset, $Re_\lambda$ is presented by a color plot on top of the relation between $C_\varepsilon$ and $\mu$. The red line indicates a commonly accepted value for $\mu$ for homogeneous isotropic turbulence~\cite{arneodo1996structure}, while the black line corresponds to a least-square fit. The symbols and corresponding configurations are shown in figure~\ref{mu_reynolds_num} from the main manuscript and explained in table~\ref{PhD measurements in LEGI 2023}. $N_v$ indicates the number of velocity time series shown in this plot.}
    \label{c_epsi_e_mu_shape_factor_all_reynolds}
\end{figure}

\begin{figure}
    \centering    \includegraphics[width=\linewidth]{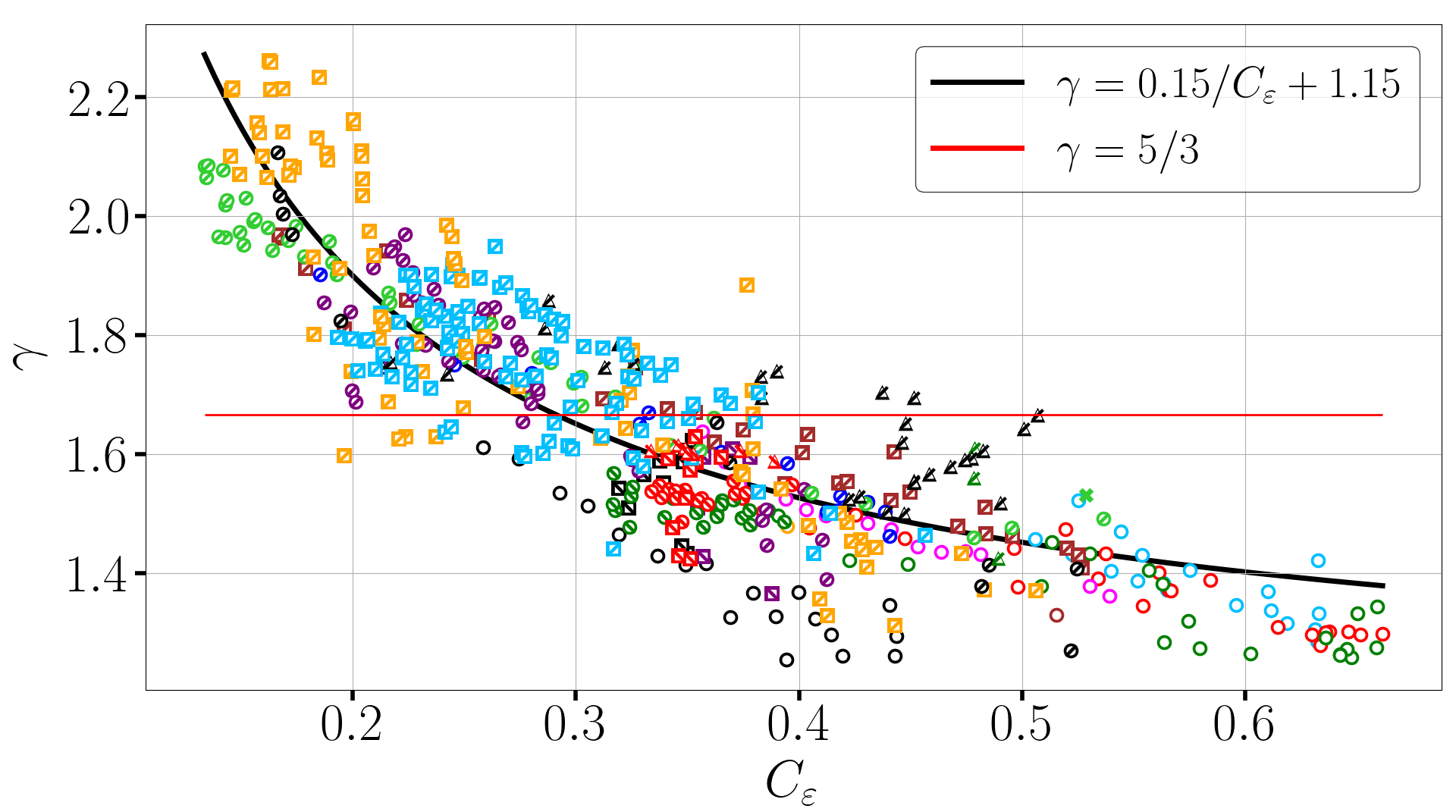} 
    \caption{\Vsix{$\gamma$ as a function of $C_\varepsilon$ for the less restricted data. The red line indicates the commonly accepted value for $\gamma$ for HIT~\cite{sreenivasan1995universality}, while the black line corresponds to a least-square fit. The symbols and corresponding configurations are shown in figure~\ref{mu_reynolds_num} from the main manuscript and explained in table~\ref{PhD measurements in LEGI 2023}.}}
    \label{gamma_c_epsi_all}
\end{figure}

\begin{figure}[h!] 
    \centering    \includegraphics[width=\linewidth]{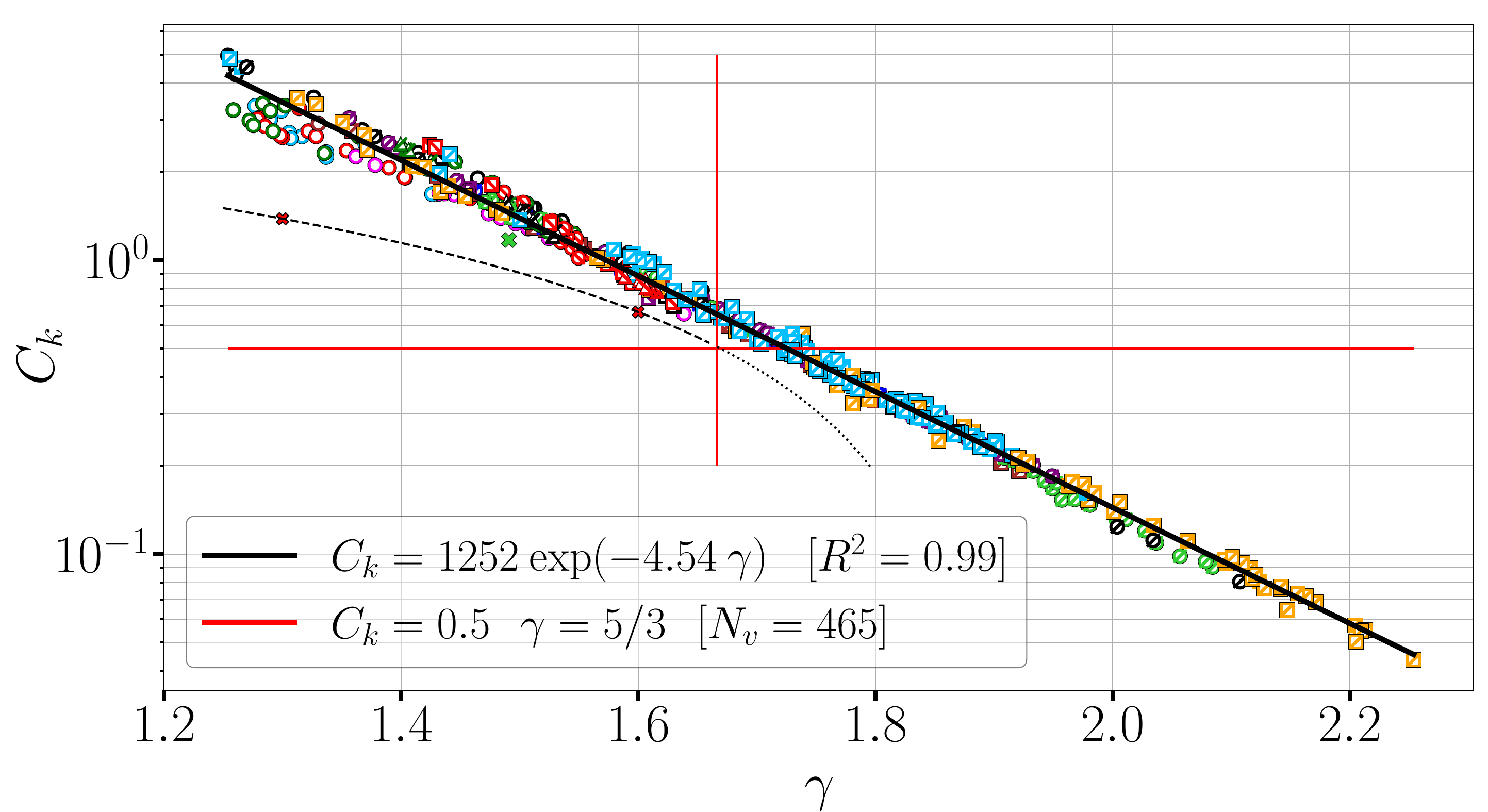} 
    \caption{$C_k$ over $\gamma$ for the less restricted dataset. The red line indicates a commonly accepted values for $C_k$ and $\gamma$ for homogeneous isotropic turbulence~\cite{sreenivasan1995universality}, while the black curve corresponds to a least-square fit ($R^2=0.99$). The dashed black curve shows the linear relationship between $C_k$ and $\gamma$ suggested by Mydlarski and Warhaft~\cite{mydlarski1996onset} until $\gamma = 5/3$. The two red crosses indicate the bounds of their measurement range. The dotted black curve represents the same linear relation for values beyond $\gamma = 5/3$. The symbols and corresponding configurations are shown in figure~\ref{mu_reynolds_num} from the main manuscript and explained in table~\ref{PhD measurements in LEGI 2023}. $N_v$ indicates the number of velocity time series shown in this plot.}
    \label{C_k_gamma_myd}
\end{figure}

\Vsix{In the following, we present results when the restrictions are softened with respect to the results presented the main manuscript.} In this case, the velocity time series that are considered, only have to fulfill the following conditions:
\noindent a) $TI$ decreases in streamwise direction,
\noindent b) the large scale increments have an almost Gaussian distribution ($\Lambda_{0}^{2} < 0.025$) and
\noindent c) the slope of the energy spectrum within the inertial range is smaller than $-1.25$, or in other words, is sufficiently steep.
\noindent After this conditioning, 466 velocity time series remained for the analysis. For the less restricted data, the acquisition time represents at least $8 \times10^3$ integral time scales. Figures~\ref{c_epsi_e_mu_shape_factor_all}-\ref{C_k_gamma_myd} show the results from the main manuscript when only low restrictions are applied. It can be seen that the overall relations stay the same while the spreading becomes larger. \Vsix{In addition, figure~\ref{C_k_gamma_myd} presents the relationship between $C_k$ and $\gamma$ suggested by Mydlarski and Warhaft~\cite{mydlarski1996onset}.}

\section*{Independence of results on the shape factor uncertainty}
  
\begin{figure}
    \centering    \includegraphics[width=\linewidth]{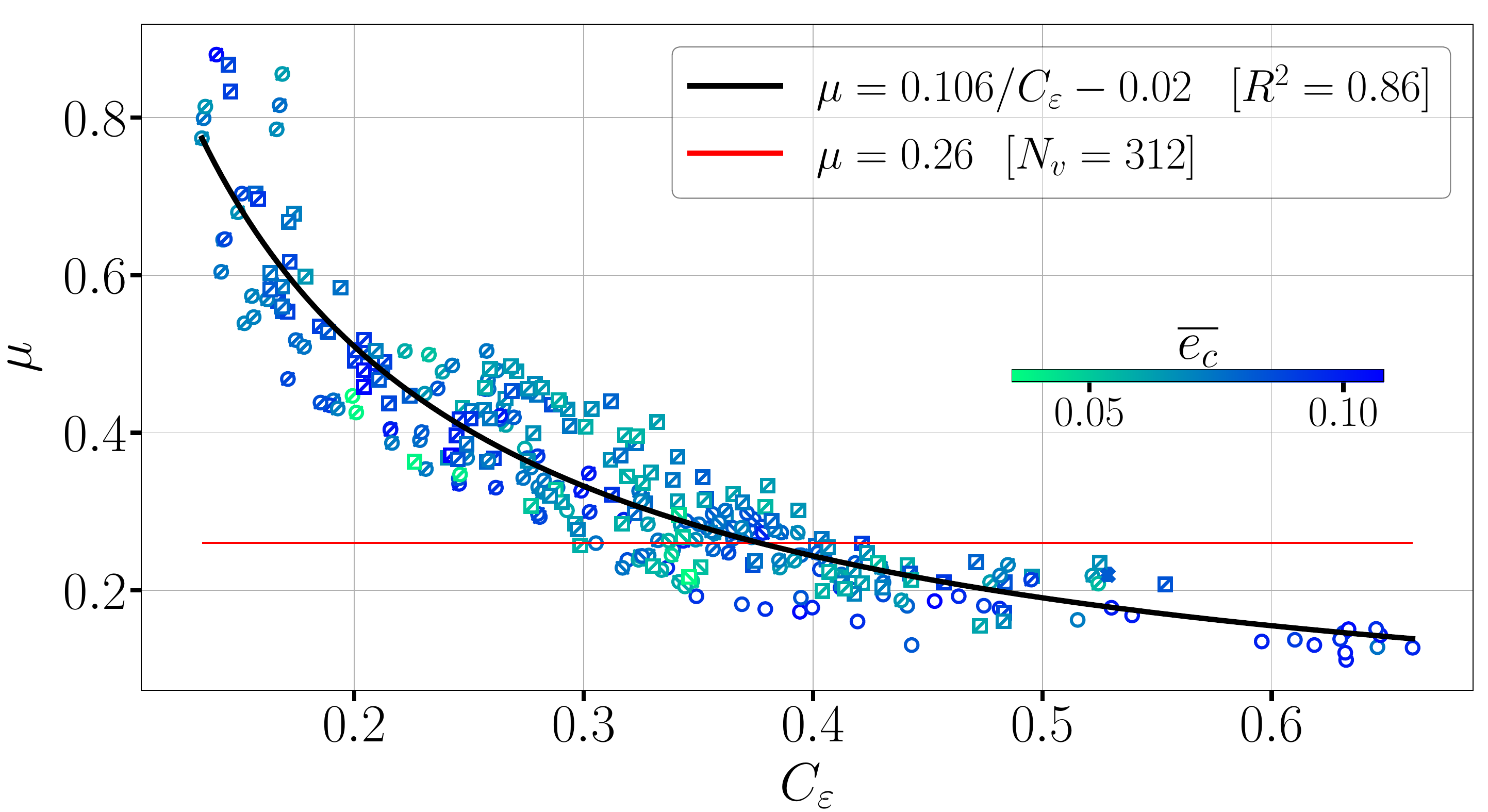} 
    \caption{$\mu$ as a function of $C_\varepsilon$ for the highly restricted dataset, where the red line indicates a commonly accepted value for $\mu$ for homogeneous isotropic turbulence~\cite{arneodo1996structure} and the black line corresponds to a least-square fit. The color demonstrates $\overline{e_c}$. The symbols and corresponding configurations are shown and explained in figure~\ref{mu_reynolds_num} and table~\ref{PhD measurements in LEGI 2023} in the appendix~\cite{SM}. $N_v$ indicates the number of velocity time series shown in this plot.}
    \label{castaing_mu_c_epsi_e}
\end{figure}

Figure~\ref{castaing_mu_c_epsi_e} presents $\overline{e_c}$ for each velocity time series as a color plot over the relation between $\mu$ and $C_\varepsilon$. 

\section*{Relation between $C_k$ and $Re_\lambda$}
  
\begin{figure}
    \centering    \includegraphics[width=\linewidth]{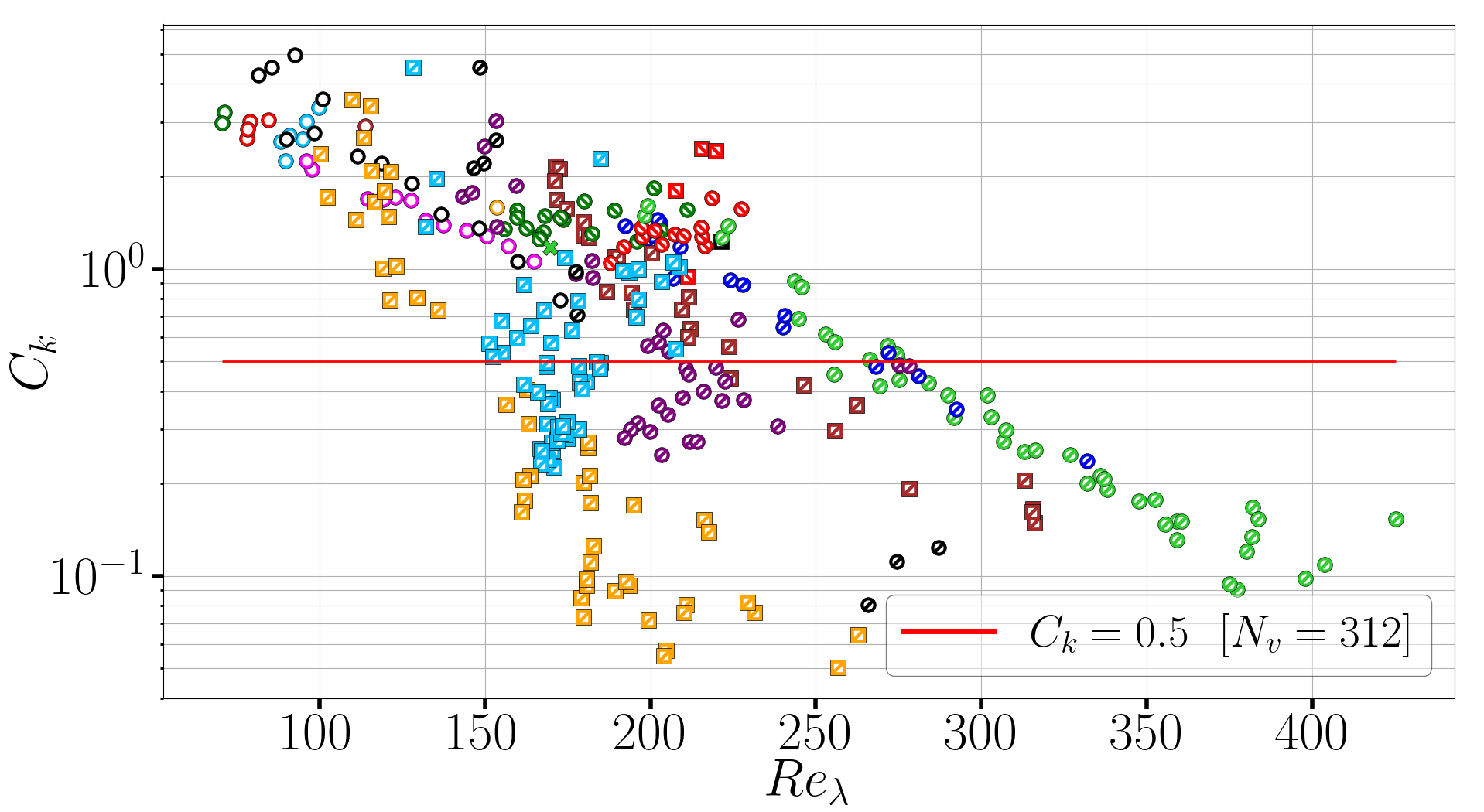} 
    \caption{\Vsix{$C_k$ as a function of $Re_\lambda$ for the highly restricted dataset, where the red line indicates a commonly accepted value for $C_k$ for HIT~\cite{sreenivasan1995universality}. The symbols and corresponding configurations are shown and explained in figure~\ref{mu_reynolds_num} and table~\ref{PhD measurements in LEGI 2023} in the appendix~\cite{SM}. $N_v$ indicates the number of velocity time series shown in this plot.}}
    \label{ck_taylor_reynolds_num}
\end{figure}

Figure~\ref{ck_taylor_reynolds_num} presents $C_k$ for each velocity time series as function of $Re_\lambda$. 

\section*{Relation between $\mu$ and $L/\eta$}

\begin{figure}[h!] 
    \centering    \includegraphics[width=\linewidth]{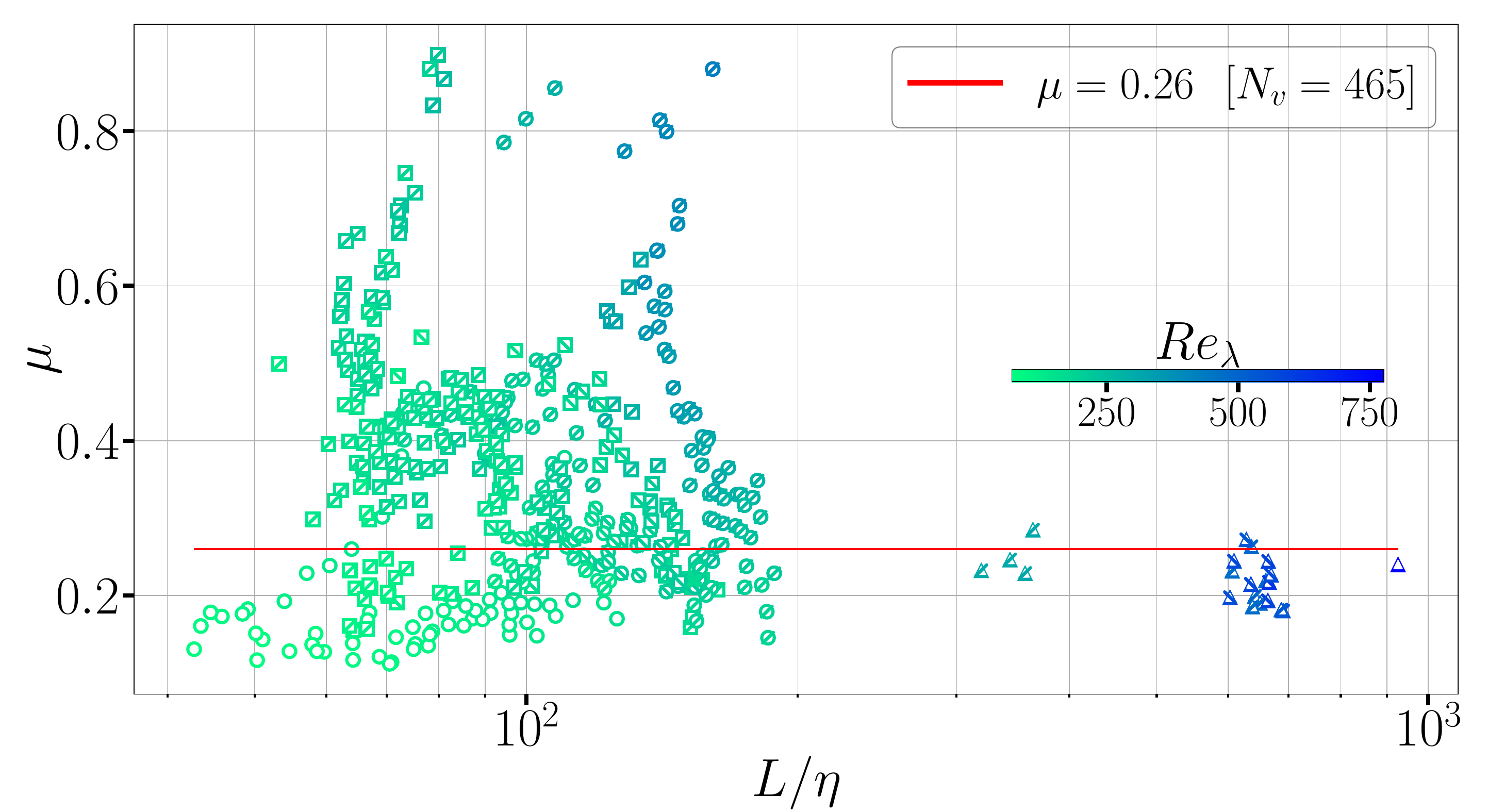} 
    \caption{$\mu$ as a function of $L/\eta$ for the highly restricted data. The red line indicates a commonly accepted value for $\mu$ for homogeneous isotropic turbulence~\cite{arneodo1996structure}. Additionally, $Re_\lambda$ is indicated by color. $N_v$ indicates the number of velocity time series shown in this plot.}
    \label{mu_L_div_eta}
\end{figure}

\Vsix{Figure~\ref{mu_L_div_eta} presents the intermittency parameter as a function of the extent of the inertial range.} 

\section*{Distribution of the skewness of velocity time signals} 

\begin{figure}
    \centering    \includegraphics[width=\linewidth]{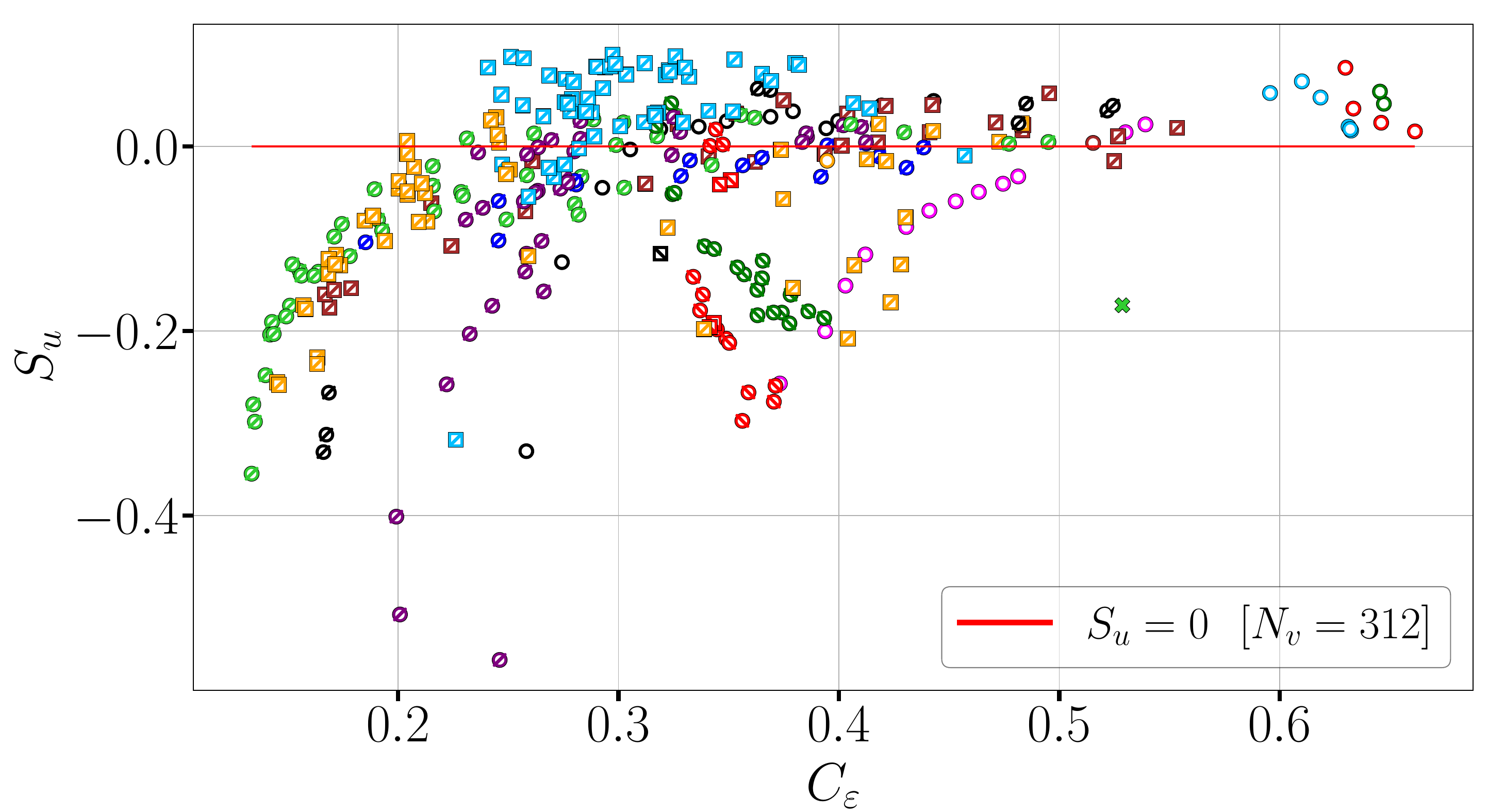} 
    \caption{Skewness of the velocity time signal $S_u$ over $C_\varepsilon$ for the highly restricted dataset. The red line divides the data in positive and negative skewness. The symbols and corresponding configurations are shown in figure~\ref{mu_reynolds_num} from the main manuscript and explained in table~\ref{PhD measurements in LEGI 2023}. $N_v$ indicates the number of velocity time series shown in this plot.}
    \label{skewness_cepsi_e}
\end{figure}

Figure~\ref{skewness_cepsi_e} shows the distribution of the skewness $S_u$ of the velocity time series over $C_\varepsilon$ for the highly restricted dataset.

\section*{Relation between $C_k$ and $C_\varepsilon$}

\begin{figure}[h!] 
    \centering    \includegraphics[width=\linewidth]{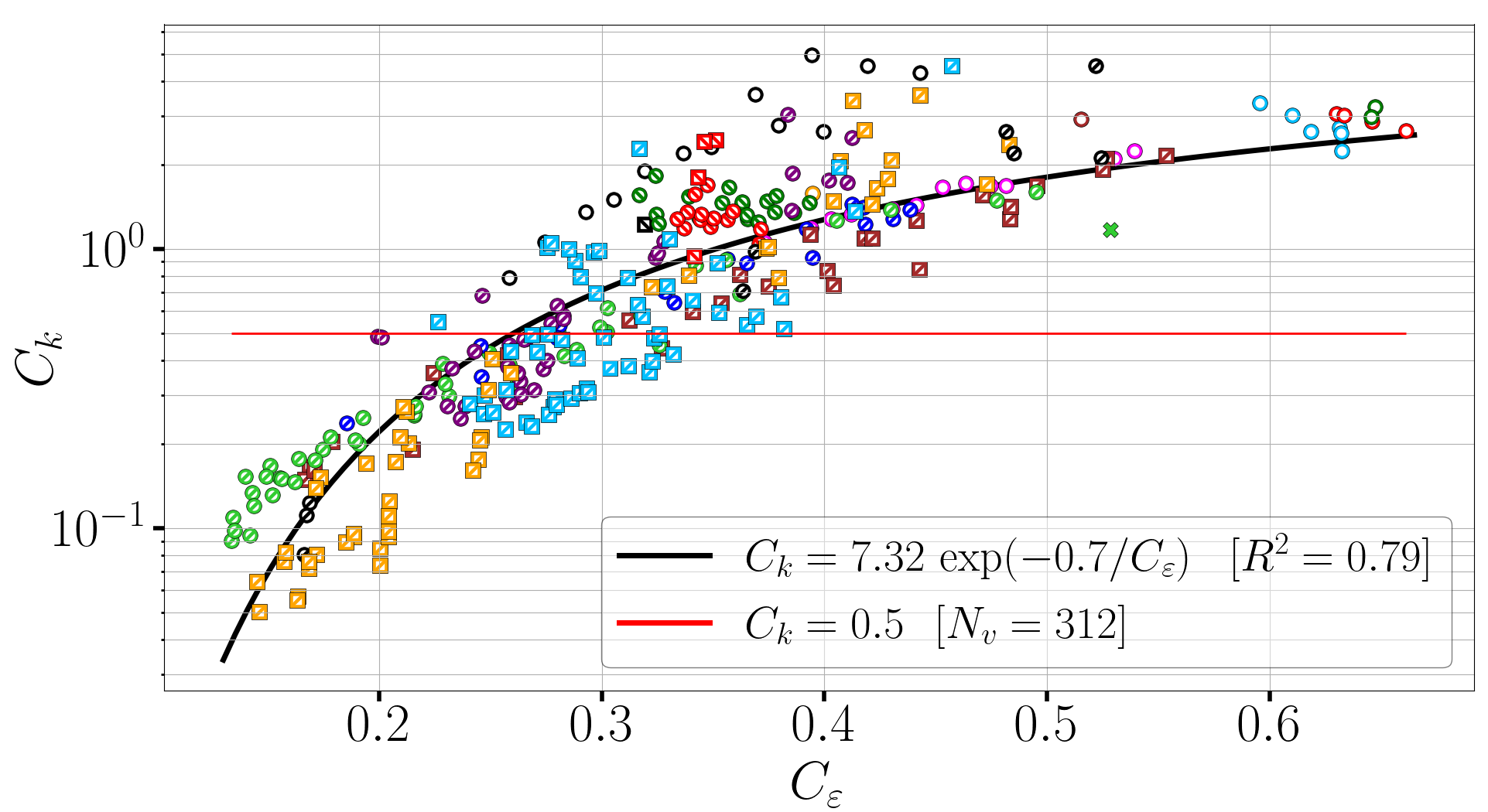} 
    \caption{$C_k$ as a function of $C_\varepsilon$ for the highly restricted data. The red line indicates the commonly accepted value for $C_k$ for HIT~\cite{sreenivasan1995universality}, while the black line corresponds to a least-square fit with $R^2$ being $0.79$. Once more, the symbols and corresponding configurations are shown and explained in figure~\ref{mu_reynolds_num} in the main manuscript and table~\ref{PhD measurements in LEGI 2023}. $N_v$ indicates the number of velocity time series shown in this plot.}
    \label{ck_c_epsi}
\end{figure}

Additionally to figure~\ref{ck_c_epsi} in the main manuscript, the relation between $C_k$ as a function of $C_\varepsilon$ is presented for the highly restricted data.

\clearpage

\section*{ACKNOWLEDGMENTS}

\noindent This work was supported by the LabEx Tec21 under Grant Investissements d’Avenir – grant agreement no. ANR-11-LABX-0030; and the Hanse-Wissenschaftskolleg (HWK Institute for Advanced Study, Delmenhorst, Germany) under a fellowship assigned to MO.

\noindent We thank Christophe Penisson, Muriel Lagauzere, Stephane Pioz-Marchand and Sylvain Dauge for helping with the experiment,
\noindent Valentin Groß for proof-reading,
\noindent Olivier De Marchi and Sebastian Bergemann for the IT support,
\noindent Thomas Messmer, Lars Neuhaus and Daniela Moreno for validating parts of the used code and
\noindent Michael Hölling, Ingrid Neunaber, Jan Friedrich and Christos Vassilicos for fruitful discussion. 
`
\section*{Author declarations}

The authors have no conflicts to disclose

\section*{DATA AVAILABILITY STATEMENT}
The data that support the findings of
this study are available from the
corresponding author upon reasonable
request.

\bibliographystyle{unsrt}
\bibliography{bib_tex_felix}

\end{document}